\documentclass[sigconf]{acmart}
\AtBeginDocument{%
  }

\acmISBN{978-1-4503-XXXX-X/2018/06}
\usepackage{xcolor}
\usepackage{multirow}
\usepackage{booktabs}
\usepackage{hyperref}

\copyrightyear{2026}
\acmYear{2026}
\setcopyright{cc}
\setcctype{by}
\acmConference[CHI '26]{Proceedings of the 2026 CHI Conference on Human Factors in Computing Systems}{April 13--17, 2026}{Barcelona, Spain}
\acmBooktitle{Proceedings of the 2026 CHI Conference on Human Factors in Computing Systems (CHI '26), April 13--17, 2026, Barcelona, Spain}
\acmPrice{}
\acmDOI{10.1145/3772318.3790812}
\acmISBN{979-8-4007-2278-3/2026/04}

\begin{document}

\title{\textcolor{black}{Sometimes You Need Facts, and Sometimes a Hug}: Understanding Older Adults’ Preferences for Explanations in LLM-Based Conversational AI Systems}

\author{Niharika Mathur}
\affiliation{%
  \department{School of Interactive Computing}  
  \institution{Georgia Institute of Technology}
  \city{Atlanta}
  \state{Georgia}
  \country{USA}
}
\email{nmathur35@gatech.edu}

\author{Tamara Zubatiy}
\affiliation{%
  \institution{Northeastern University}
  \city{Boston}
  \state{Massachusetts}
  \country{USA}
}
\email{t.zubatiy@northeastern.edu}

  \author{Agata Rozga}
\affiliation{%
  \department{School of Interactive Computing}
  \institution{Georgia Institute of Technology}
  \city{Atlanta}
  \state{Georgia}
  \country{USA}
}
\email{agata@gatech.edu}

  \author{Jodi Forlizzi}
\affiliation{%
  \department{Human-Computer Interaction Institute}
  \institution{Carnegie Mellon University}
  \city{Pittsburgh}
  \state{Pennsylvania}
  \country{USA}
}
\email{forlizzi@cs.cmu.edu}

  \author{Elizabeth D Mynatt}
\affiliation{%
  \department{Khoury College of Computer Sciences}
  \institution{Northeastern University}
  \city{Boston}
  \state{Massachusetts}
  \country{USA}
}
\email{e.mynatt@northeastern.edu}



\renewcommand{\shorttitle}{Understanding Older Adults’ Preferences for Explanations in LLM-Based Conversational AI Systems}
\renewcommand{\shortauthors}{Mathur et al.}

\begin{abstract}

Designing Conversational AI systems to support older adults requires these systems to explain their behavior in ways that align with older adults’ preferences and context. While prior work has emphasized the importance of AI explainability in building user trust, relatively little is known about older adults’ requirements and perceptions of AI-generated explanations. To address this gap, we conducted an exploratory Speed Dating study with 23 older adults to understand their responses to contextually grounded AI explanations. Our findings reveal the highly context-dependent nature of explanations, shaped by conversational cues such as the content, tone, and framing of explanation. We also found that explanations are often interpreted as interactive, multi-turn conversational exchanges with the AI, and can be helpful in calibrating urgency, guiding actionability, and providing insights into older adults’ daily lives for their family members. We conclude by discussing implications for designing context-sensitive and personalized explanations in Conversational AI systems.
\end{abstract}



\maketitle

\section{Introduction}

Conversational AI systems are increasingly becoming part of older adults’ lives and are emerging as promising tools to support them. As of 2025, 55\% older adults report having used Conversational AI technologies such as voice or text-based assistants (Siri, Alexa, ChatGPT, etc.) \cite{UMichAI2023}. Aging in place, defined as \textit{home-based care}, enables older adults to stay at home as they grow older \cite{caldeira2017senior, soubutts2021aging}. It has traditionally been viewed as an effective approach to support healthy aging due to benefits of familiarity and independence \cite{wiles2012meaning}, a need that has become more pronounced since the pandemic \cite{holaday2022loneliness}. For older adults navigating cognitive changes, like those associated with Mild Cognitive Impairment (MCI), maintaining a sense of autonomy and functional independence while aging remains deeply important, as their needs and abilities evolve and change over time \cite{fang2017informing}. The growing adoption of Conversational AI systems among older adults can be attributed to their interactive conversational modality \cite{wulf2014hands, arnold2024does}, relatively easier learning curve \cite{desai2023ok, jakob2025adapting}, increased anthropomorphization \cite{chin2021being, pradhan2025no} and efficient information access in smart home-enabled environments \cite{kim2021exploring, ma2023adoption, el2018towards}. More recently, Large Language Models (LLMs) have shown potential in enabling more flexible, responsive, and context-aware interactions by expanding the range of conversational abilities of such systems \cite{yang2024talk2care, shahid2025exploring}.  

However, the increased conversational depth of LLM-based systems also introduces parallel challenges. One such challenge concerning older adults’ trust and confidence is related to the AI’s \textbf{explainability}, or lack thereof \cite{mathur2023did, gleaton2023understanding, enam2025artificial}. In theory, AI explanations are system outputs that reveal the inner mechanisms of AI systems in order to provide reasoning or justification for their responses. Concerns around explainability are central to the area called Explainable AI (XAI), which focuses on generating intelligible justifications for AI behavior. For example, when an AI system reminds an older adult to take a medication, a natural user response might be, \textit{“Why are you reminding me now?”}, with the expectation for the AI to explain why it generated the reminder. Such a response may stem either from a perceived mismatch between their own memory and the AI's timing \cite{srinivasan2021explanation}, or from a broader questioning of the AI’s reliability and trustworthiness \cite{de2017people}. Research on AI explanations first emerged alongside early intelligent systems. Yet, most XAI approaches today continue to prioritize transparency for AI engineers and developers, rather than for everyday end-users interacting with these systems \cite{miller2019explanation, turri2024transparency}. However, as AI systems increasingly begin to mediate important decisions in interpersonal settings like the home, explanations of AI responses become crucial factors influencing people’s trust and reliance on them. 

\textcolor{black}{For older adults, appropriate explanations of AI behavior can be particularly helpful for critically reflecting on AI-generated outputs. In increasingly technology-rich home environments, explanations can help them make sense of why an AI is acting in a particular way, reflect on how it interprets their routines and exercise agency in responding or overriding its suggestions  \cite{mathur2023did, langley2017explainable, mynatt2025technology}. Additionally, as social supports and family communication channels evolve, explanations can also function as communicative bridges between older adults, their technologies, and other people in their lives, thus necessitating explanation strategies that are attuned to both individual and collaborative needs.} 

While prior work has emphasized the importance of AI explainability in building user trust, relatively little is known about how older adults perceive or respond to explanations. In this paper, we report findings from an exploratory investigation into older adults’ preferences and perceptions towards AI-generated explanations. We focus in particular on explanations that are drawn from diverse sources of information present in their homes. To address this, we conducted a Speed Dating study \cite{zimmerman2017speed} with 23 older adults, using storyboard-based vignettes across two assistive contexts: \textbf{routine reminders} and \textbf{emergency alerts}. Participants encountered an AI system (depicted as a fictional robot in the storyboards) that provided them a series of explanations. Participants then reflected on their needs and preferences for explanations, discussing thoughts around explanatory functions, expectations, and goals. 

Our findings reveal a strong influence of task context on older adults’ explanation preferences, with explanatory value being evaluated based on a scenario’s perceived risk. We also observed distinct differences in responses to explanations that referenced prior conversations with the AI versus those grounded in real-time observations from a user’s surroundings, such as activity sensors or wearables. \textcolor{black}{We also observed instances where some partners described their vision for AI explanations as helpful “windows” into each other's daily routines, reflecting need for AI explanations to attune to the shared and coordinated nature of household life among older adults. Such interpretations emerged naturally as a result of spousal partners sharing co-living responsibilities; however, in this paper, we do not intend to analyze caregiving relationships specifically.}

\subsection{Contributions}
This work offers the following contributions to the larger HCI research community:

\begin{itemize}
    \item Empirical insights into older adults’ preferences for AI explanations, highlighting how requirements shift across different assistive contexts. This work also underscores the need for Conversational AI systems to better align with older adults’ goals by adapting explanatory tone, content, and conversational framing to the context and users’ emotional states at the time of interaction.
    \item Examination of how older adults interpret AI explanations that are drawn from different sources of information, such as prior conversational history or real-time environmental data, and articulation of how these sources influence perceptions of trustworthiness, clarity, and usefulness in explanations. 
    \item Insights on the multifaceted roles and functions of AI explanations beyond technical transparency, showing how older adults view them as part of an interactive, multi-turn conversational exchange, capable of calibrating urgency, guiding actionability, and serving to enhance \textcolor{black}{awareness of their partner's daily routines.} 
\end{itemize}

\section{Related Work and Background}

In this section, we discuss older adults’ use of Conversational AI systems, focusing on expectations for transparency and explainability. We then explore how LLMs present emerging opportunities to design personalized explanations and the need to examine explanatory preferences in such systems. We conclude Section 2.3 with our research questions based on a review of related work. 

\subsection{Older adults’ Use of Conversational AI Systems}
The design of AI-driven assistive technology in supporting older adults is a growing area of focus within HCI, including those that have explored the potential of Conversational AI systems \cite{sengupta2020challenges, bokolo2025examining, corbett2021voice} or more specifically, Voice-First Ambient Interfaces (VFAIs) \cite{cuadra2023designing, so2024they, kokorelias2024longitudinal, liu2023older}. These systems are either embedded within smartphones \cite{kurniawan2008older, west2019s, dixon2022mobile}, as standalone devices or smart speakers \cite{park2022impact, pradhan2020use, brewer2022empirical, chen2021understanding, brewer2022if}, or physically instantiated as assistive robots providing holistic support in homes \cite{pino2015we, chang2024dynamic, chang2025unremarkable, getson2021socially}.

Advances in ubiquitous and pervasive computing have spurred investigations into sensing-based approaches for effective data capture in smart home environments supporting aging in place \cite{mynatt2001digital, cozza2017ubiquitous, ma2023adoption, el2018towards}, with studies highlighting older adults’ willingness to engage with sensing-enabled support for better assistance \cite{townsend2011privacy, pirzada2022ethics}. This line of research points to the potential of routine-based support \cite{davidoff2010routine, rowan2005digital, consolvo2006design}, and context-awareness \cite{aloulou2023personalized, mihailidis2002context} in delivering effective reminders (e.g., for medication \cite{lee2015sensor, mathur2022collaborative}), safety alerts (e.g., fall detection \cite{wu2018understanding, ejupi2016kinect, mohan2024artificial}), companion-like support via social and conversational functionalities \cite{el2020multimodal}. \textcolor{black}{These works widely characterize home-based assistive interactions along a continuum from routine, low-risk support to higher-risk safety monitoring, a framing that also informed our selection of reminders and alerts as representative contexts for exploring explanation preferences (further discussed in Methods).}

While these works demonstrate a promising use case for contextually rich data in supporting assistance through Conversational AI, several challenges persist. Longitudinal evaluations of such systems have identified practical issues that arise when older adults interact with them over time. Zubatiy et al. highlight the difficulty of managing complex interactions that extend beyond simple, task-based requests and require AI systems to engage users proactively \cite{zubatiy2023don}. In separate studies, Mathur et al. and Gleaton find that older adults frequently questioned AI systems and expected useful explanations \cite{mathur2023did, gleaton2023understanding}. Concerning privacy disclosures, Enam et al. identify the “black-box” nature of chatbots as a key factor in older adults’ reluctance to engage with them \cite{enam2025artificial}. Most existing explainable AI applications focus on narrow, low-level explanations of individual decisions rather than providing the explanatory depth humans require for trust and acceptance \cite{dazeley2021levels, sokol2018conversational}. This need for effective explainability, among other factors, has shown to influence AI adoption due to uncertainties around the system’s intent, decision-making, and trustworthiness \cite{bhushan2024explainable, ferrario2022explainability, weitz2019you, rai2020explainable}. 

Broadly, explanations can be understood as answers to \textit{why} questions that aim to help users interpret the AI’s underlying reasoning and understand sources for its outputs \cite{kim2025fostering, xu2019explainable, arrieta2020explainable}. Srinivasan and Chander find that explanations can serve as key cognitive functions that help users both recall and remember prior interactions and develop a more coherent understanding of the AI’s behavior \cite{srinivasan2021explanation}. For older adults experiencing memory-related challenges due to MCI or dementia, this perspective further motivates the importance of studying effective explanation design. In such instances, appropriately designed explanations have the potential to improve understanding, support memory retention, and facilitate sustained interactions with AI systems.

Our work is motivated by this growing body of research examining explainability of AI systems designed to support older adults. The timeliness of this research is underscored by recent advances in AI reasoning capabilities, particularly those enabled by LLMs, showing potential for LLM-based systems to provide more detailed explanations to users \cite{zytek2024llms, ehsan2025new, cambria2024xai}.

\subsection{Opportunities and Challenges with LLM-generated Explanations}

LLMs now enable the generation of conversational and flexible natural language explanations, expanding on traditional explainability approaches that largely relied on post-hoc visualizations  \cite{bhatt2020explainable, miller2017explainable}. These explanations can also draw on a user’s prior conversational history and be expressed in natural language \cite{sebin2024exploring, zhao2023survey}. 

However, while transformative, LLMs are not a panacea for the nuances of explainability in AI systems, particularly in sensitive contexts \cite{suh2025don, ferrario2024addressing, yang2024mentallama}. Emerging research has identified several limitations, where known LLM usability issues affect their ability to deliver meaningful explanations. These issues include opacity \cite{manche2022explaining}, hallucination \cite{perkovic2024hallucinations, orgad2024llms}, sycophantic tendencies \cite{sun2025friendly, carro2024flattering}, and lack of transparency in training data \cite{ehsan2024human}, all of which undermine the reliability, consistency, and trustworthiness of LLM-generated explanations. 

Furthermore, Ehsan et al. highlight the challenge of translating multi-billion-parameter LLMs into meaningful explanations \cite{ehsan2024human}. In a large-scale experiment, Kim et al. observed that LLMs’ use of uncertainty expressions influenced trust and reduced overreliance, underscoring the difficulty of fine-tuning language to calibrate user understanding \cite{kim2024m}. Suh et al. attribute this overreliance to the friendly, conversational tone of LLMs, arguing that users trust explanations \textit{“simply because they exist”}, regardless of accuracy  \cite{suh2025don}. Similarly, He et al. find that the interactive nature of LLMs contributed to creating a false perception of trust \cite{he2025conversational}. 
Herrera describes the complexity of designing truthful LLM explanations, i.e., how accurately the content aligns with external facts. To address this, they propose mechanisms focused on \textit{functional intelligibility}, enabling users to ask questions and receive tailored explanations \cite{herrera2025making}. This body of work calls for careful evaluation and user-centered studies before deploying LLM-generated explanations at scale to recognize their limitations.

\subsection{Towards Human-Centered AI Explanations}

In addition to LLM-specific limitations, the broader algorithm-centric focus in explainable AI research has also faced scrutiny in recent years. HCI researchers, specifically, have called for a deeper understanding of users’ explainability needs in everyday contexts  \cite{hong2020human, brennen2020people, miller2024explanation}. This perspective treats explanations as communicative responses that must be grounded in the user’s environment to enhance contextual understanding \cite{ehsan2022human, ribera2019can}. 

Prior work shows that explanation needs vary significantly across user groups, requiring explanation strategies that are sensitive to users’ goals and expectations \cite{ridley2025human, xu2023hcai, liao2020questioning, ehsan2023human}. For example, while developers may seek technical explanations for debugging and model comprehension, non-technical users often benefit from conversational explanations embedded in their interaction goals \cite{mansi2025legally, huang2024research, shandilya2022understanding}. 

To address these concerns, the emerging field of Human-Centered Explainable AI (HCXAI) calls for empirical examinations of explainability perceptions that prioritize user perspectives \cite{rong2023towards, ehsan2021expanding}. For example, Ehsan and Riedl, through example of a judgment-based task, advocate for explanations to include environmental information outside the AI to enhance understanding of the user’s context \cite{ehsan2020human}. Additionally, design-based approaches, such as research-through-design, have also been proposed as exploratory steps prior to technical deployment  \cite{mathur2025research}. Ferreira and Monteiro propose a useful \textit{“who”}, \textit{“why”}, and \textit{“what”}  framework to guide human-centered evaluations of AI explanations, emphasizing: (1) intended audience of the explanation (\textit{who}), (2) explanation’s function or purpose (\textit{why}), and (3) domain and context of interaction (\textit{what}) \cite{ferreira2020people}. Among these, the first and third dimensions, focused on the explanation recipient's identity and context, are particularly relevant to our goals. Using an open-ended, exploratory design approach, we foreground the understanding of explanatory \textcolor{black}{perceptions} as a foundation for  our work and pose our first research question: \textcolor{black}{\textbf{RQ1: }\textit{How do older adults perceive the roles and value of AI explanations in supporting their decision-making across everyday interactions?}} 

\subsubsection{AI Explanations for Aging in Place}

Aging in place represents a unique intersection of cognitive, social, and environmental factors that shape and influence how older adults interact with assistive technologies \cite{bagnall2006older, zubatiy2023distributed, kaliappan2024exploring}. In such settings, the consequences of appropriately understanding (or misunderstanding) the AI’s behavior can be particularly high. Prior studies have shown that older adults often draw on elements of their sociotechnical environment during interactions \cite{pradhan2021entanglement, sin2020empirically, jacelon2013older}, with broader sociocultural influences, such as attitudes toward privacy, also playing a role \cite{you2025sociocultural, fritz2015influence}. Pradhan et al. describe this sociotechnical landscape as a network of physical and digital artifacts that mediate older adults’ engagement with AI technologies \cite{pradhan2023towards, pradhan2018accessibility}. Familiar household objects, such as paper calendars, have similarly been shown to influence the mental models older adults form, shaping how they interpret and respond to AI behaviors \cite{zubatiy2023distributed}.

Along these lines, there is a growing interest in studying how information from sociotechnical contexts can inform explanation design. In a scoping review, Baruah and Organero highlight the potential of incorporating domain knowledge, such as user preferences and past queries, into explanations to enhance usefulness \cite{baruah2024brief}. Alizadeh et al. characterize this sociotechnical awareness in explanations as central to a \textit{holistic user experience} \cite{alizadeh2023user}. Drawing from activity theory, Kofod-Petersen and Cassens propose grounding explanations in users’ ongoing activities and interaction patterns in ambient smart home environments \cite{kofod2007explanations}. This is further operationalized in \cite{das2023explainable} using activity recognition frameworks to generate natural language explanations, with 83\% of participants preferring contextually grounded versions over traditional activity labels.  Here, the dual role of environmental context is emphasized: \textit{first}, in shaping how the system reasons about a situation; \textit{second}, in how it communicates that reasoning to the user. Our research goals align most closely with the latter, i.e., how explanations bridge the gap between system reasoning and user expectations. Furthermore, Mathur et al. highlight the potential of incorporating elements of users’ sociotechnical environments into explanations, drawing on longitudinal fieldwork with older adults  \cite{mathur2024categorizing}. Based on their analysis, they propose a categorization of four key information sources that AI systems can leverage when generating explanations in smart home environments, discussed more in the next section \cite{mathur2024categorizing}. Inspired by their structured approach, we use this categorization of information sources to design the explanation prompts in our study. 

To our knowledge, our study is the first empirical investigation of older adults’ perceptions and preferences for AI-generated explanations. Here, we articulate our second research question: \textcolor{black}{\textbf{RQ2:} \textit{How do older adults’ perceptions and the way they make sense of AI explanations vary by contexts of use (routine vs. urgent) and by explanation type (differentiated by information sources in the home)?}}

\section{Method}

To understand how older adults perceive AI explanations, we conducted a Speed Dating study \cite{zimmerman2017speed, davidoff2007rapidly}, composed of a series of structured User Enactments (UEs) \cite{odom2012fieldwork}. The aim of our study was to gain insights into what type of contextually grounded explanations in a smart home environment could be helpful, and what kind of explanations may seem disruptive to older adults, driven by their needs, goals, and boundaries for information access.  

Described as “fieldwork of the future”, Speed Dating is an exploratory research method that adapts the metaphor of romantic speed dating by presenting participants with brief and structured design concepts one after the other to surface their preferences, priorities, and boundaries for future technologies \cite{zimmerman2017speed}. In a Speed Dating study, participants encounter a series of storyboards with multiple design alternatives (in our case, multiple explanation possibilities), and are then prompted to critically reflect on the implications of each. The design alternatives serve as prompts to evoke further discussion and surface user needs around technological interactions.

\textcolor{black}{Our choice of the Speed Dating method for this study was driven by its low-fidelity and future-focused nature. We use this method as a means of examining user preferences for contextually-grounded AI explanations before committing to high-fidelity, more interactive prototypes or technical development. Prior design-led HCI work has shown that low-fidelity methods can help participants stay focused on the underlying values, expectations and everyday integration of emerging technologies rather than on the usability constraints of a specific artifact \cite{reig2023dreaming, walker2002high}. For older adult populations in particular, higher-fidelity prototypes can inadvertently shift attention toward the unfamiliar interface itself and risk constraining evaluation to the artifact or prototype itself rather than the underlying interaction that we aim to study \cite{vines2015age, pradhan2023towards}. Given that our goal at this stage was to explore a broad range of explanation types in order to uncover contextual preferences that can meaningfully inform subsequent technical development, we decided to use the Speed Dating method to address our research goals.}

\textcolor{black}{Speed Dating} has been widely applied in HCI to support exploratory examination across diverse areas, including social boundaries of AI agents \cite{luria2020social}, AI affiliation in collaborative caregiving \cite{chang2024dynamic}, online mental health communities \cite{stapleton2024if}, configurations of AI embodiment \cite{luria2019re}, etc.\textcolor{black}{, demonstrating its value in surfacing early-stage insights into technologies that are not yet fully realized. In line with this tradition, we use Speed Dating as an initial exploratory and generative step towards understanding older adults' perceptions of AI explanations. Questions about explanation accuracy, reliability, and handling uncertainty in interactive systems are important next directions to ensure more generalizability of findings. We discuss these technological limitations and their implications for future high-fidelity prototypes in the Limitations and Future Work section.} 

In designing our UEs, we drew on scenario-driven brainstorming to explore how contextual information in a smart home environment could be integrated into AI explanations. We enacted the scenarios through digital storyboards, as all interactions in our study occur within the home environment. We further drew methodological inspiration from Luria et al.’s incorporation of structured design alternatives within traditional user enactments \cite{luria2019re, reig2020not}. Our aim was to move beyond fully open-ended responses and to elicit older adults’ perceptions of explanations that are structured and differentiated by the source of information contained within them. Prior scenario-driven work has shown that older adults often find unconstrained “blue-sky” ideation challenging, and benefit from having structured prompts to guide ideation \cite{waycott2019designing, vines2015age}. Consequently, our structured adaptation of UEs provided an engaging and accessible experience for participants and also supported open exploration of future explanation possibilities. 

\begin{figure*}[t]
  \centering
  \includegraphics[width=0.70\textwidth]{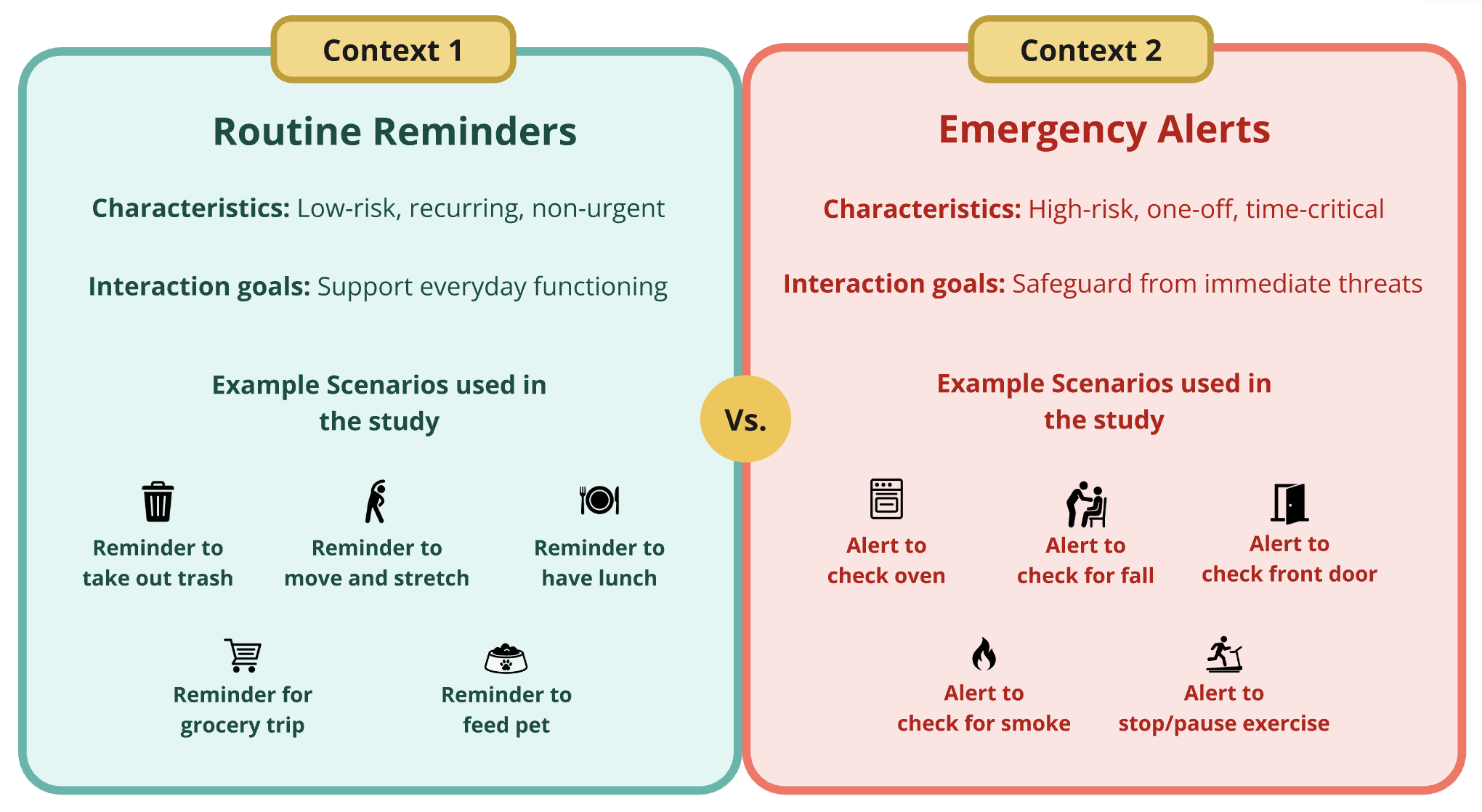}
  \caption{The two assistive contexts in the study. Context 1 represents \textbf{Routine Reminders} that are low-risk, recurring, and non-urgent. Context 2 represents \textbf{Emergency Alerts} that are high-risk, one-off, and time-critical.}
  \label{fig:fullwidth}
  \Description{This is a description of the two assistive contexts used in the study. The first context, Routine Reminders, consists of tasks that are low-risk, recurring and non-urgent. The interaction purpose in this context is to support everyday functioning. The second context, Emergency Alerts, consists of tasks that are high-risk, one-off, and time-critical. The interaction purpose in this context is to safeguard users from immediate threats.}
\end{figure*}

We structured and designed our explanations based on the categorization proposed in \cite{mathur2024categorizing}. In \cite{mathur2024categorizing}, they list and categorize the available sources of information that can be leveraged to generate AI explanations into the following four types, drawing from longitudinal fieldwork studying older adults’ interactions with AI systems in smart home-enabled environments.

\begin{enumerate}
    \item \textit{Conversational history:} includes a user's prior \textbf{interaction history} with the AI; 
    \item \textit{Environmental data:} includes real-time information from \textbf{sensor triggers} or \textbf{wearables};
    \item \textit{Activity-based inferences:} includes information and inferences about specific \textbf{user activities};
    \item \textit{Internal system logic:} includes model \textbf{confidence scores} (containing no external user data).
\end{enumerate}

In \cite{mathur2024categorizing}, the fourth category is included as a representation of a common practice in XAI of communicating AI certainty through confidence scores, \textcolor{black}{in which the model confidence score is in itself intended to be an explanation \cite{moradi2021post, poursabzi2021manipulating, le2023explaining}.} A categorization of this type is particularly valuable in home environments where the availability of rich and multimodal data about users and their routines can enable the generation of contextually grounded explanations. Adopting this categorization in our study also allowed us to evoke meaningful discussions and examine how older adults respond to explanations  containing different types of information about themselves and their environment.

\begin{figure*}
    \centering    \includegraphics[width=0.9\textwidth]{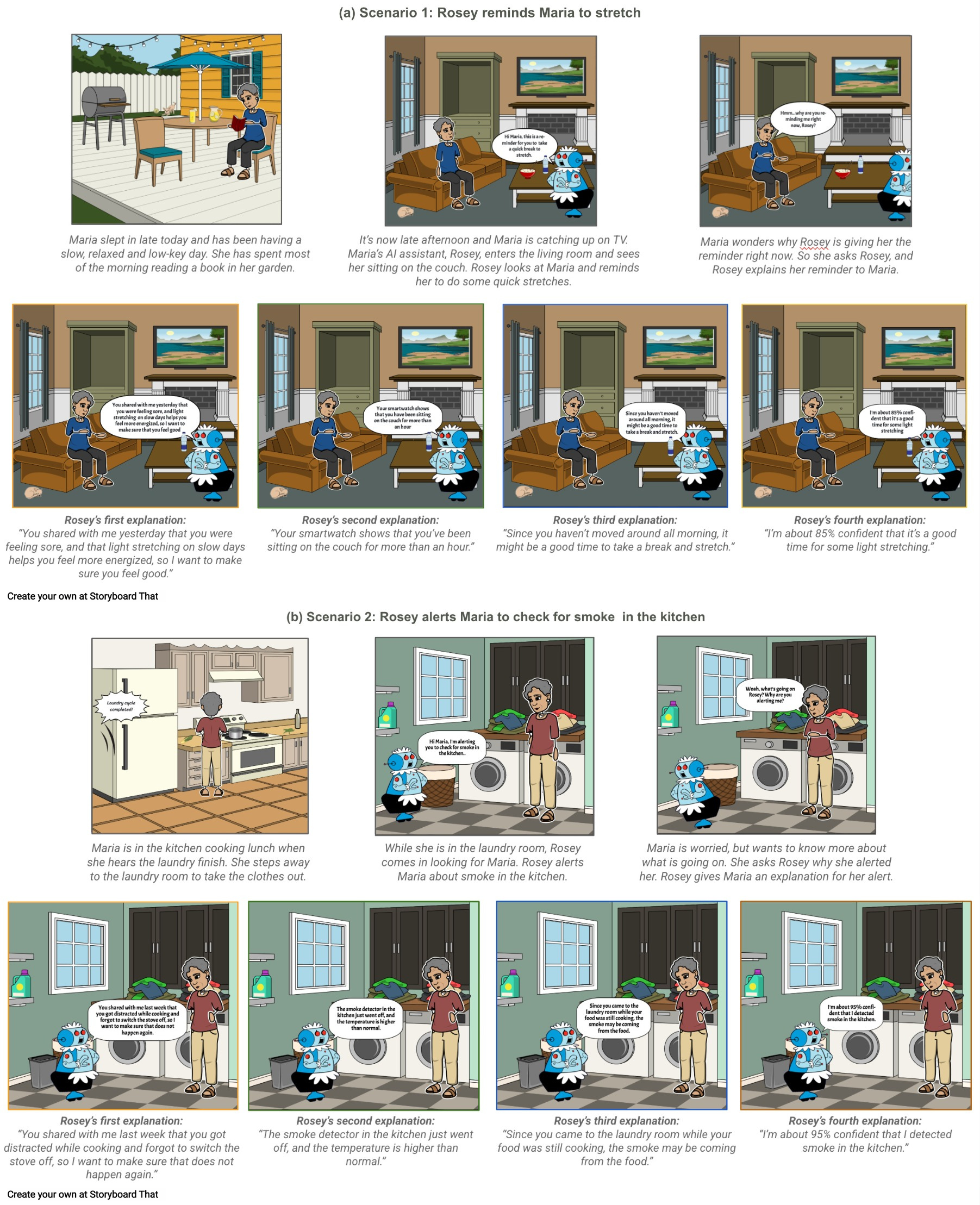}
    \caption{\textcolor{black}{Two examples from the final set of storyboards. The two scenarios depict an AI reminding an older adult to stretch during the day in (a), and an alert to check kitchen for smoke in (b). The two scenarios in (a) and (b) represent the two distinct assistive contexts in the study, i.e., \textbf{routine reminders} and \textbf{emergency alerts}, respectively. In each of the scenarios, the AI, called \textit{Rosey}, interacts with the older adult and is asked to provide an explanation. The four vignettes after the explanation request represent the four different explanation designs explored in this study, structured using the categorization in \cite{mathur2024categorizing}. This image was created using \href{https://www.storyboardthat.com/}{StoryboardThat.com}}}
    \Description{This figure illustrates two sample storyboards used in the study to evoke discussions during the Speed Dating sessions. The two scenarios presented here depict an AI system reminding an older adult to stretch during the day in (a), and an AI system providing an alert to an older adult to check their kitchen for smoke in (b). The two scenarios in (a) and (b) represent the two distinct assistive contexts in the study, i.e., routine reminders and emergency alerts, respectively. In each of the scenarios, the AI, called Rosey, interacts with the older adult and is asked to provide an explanation for her reminder in (a), and alert in (b). The four vignettes after the explanation request represent the four different explanation designs explored in this study, structured using the categorization introduced by Mathur et al. These storyboards were Created with Storyboard That.}
 \end{figure*}

\subsection{Ideation Process}
Informed by prior work, we defined two distinct assistive use contexts for our study \cite{corbett2021voice, kocaballi2020responses, chin2024like}. We define “context” here as the confluence of the physical, social, and emotional conditions surrounding an interaction. The following subsections detail our description of the two contexts and the storyboard development.

\subsubsection{Contexts of Use}
The two contexts of use in our study reflect a continuum of support older adults may require: from routine assistance to urgent and safety-critical interventions. \textcolor{black}{This choice of contexts was informed by prior work on studying smart home assistance \cite{morris2013smart, aggar2023smart, mynatt2001digital, timon2024developing}, activity-recognition systems \cite{das2023explainable}, and Conversational AI use by older adults \cite{pradhan2020use, zubatiy2021empowering, mathur2022collaborative}. Across these works, we noted that older adults often either used or expressed a preference to use AI systems to set up self-initiated reminders and safety interventions. For example, both Pradhan et al. and Zubatiy et al. report preferences to configure daily reminders and safety warnings in AI systems for aging in place \cite{pradhan2020use, zubatiy2021empowering}. Furthermore, Aggar et al. highlight that warning alerts in smart home technologies contributed substantially to older adults’ sense of safety in longitudinal deployments \cite{aggar2023smart}.} \textcolor{black}{In prior work, such reminders have also been reimagined as “check-ins,” where older adults expressed preference for AI systems that periodically touch base with them about routine activities throughout the day \cite{mathur2022collaborative}.} \textcolor{black}{Additionally, through detailed interviews with older adults, Chan et al. identify two primary contexts of relevance for AI-assisted interventions for older adults aging in place. They describe the first as a routine-based context, involving \textit{“recurring habits, preferences or practices”} that often benefit from reminders, and the second as a real-time trigger context, which include \textit{"actions that must be taken at critical moments"}, reinforcing these as meaningful and stakeholder-relevant contexts for exploring explanatory needs in our study \cite{chan2025insights}.} 

\textcolor{black}{Drawing on this body of evidence, we selected reminders and alerts to sample two distinct points along the design space of assistive AI interactions while reflecting stakeholder-driven needs. This framing allowed us to examine how explanation expectations vary across contexts that are meaningful in older adults’ lives and differentiated by their risk and time-critical characteristics.} Figure 1 summarizes the two contexts with examples. 

\vspace{1em}

\begin{figure*}
    \centering
\includegraphics[width=1\linewidth]{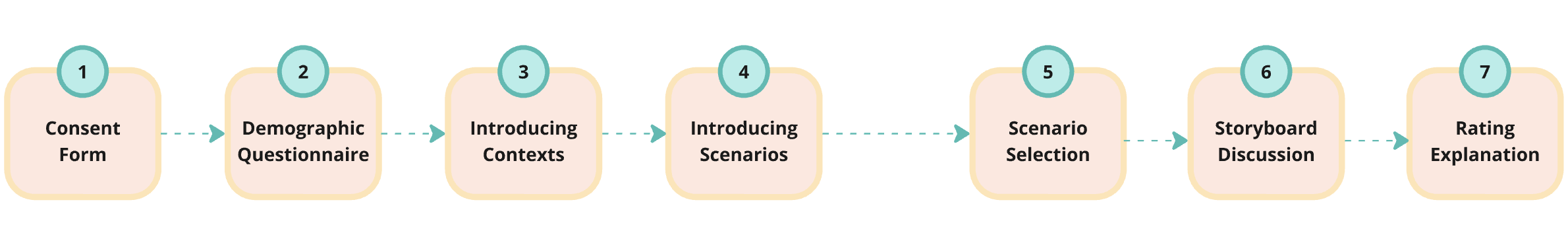}
    \caption{Study Session Flow.}
    \Description{This figure visualizes the flow in a single study session starting with consent form, demographic questionnaire, introducing contexts, introducing scenarios, scenario selection, Storyboard discussion, and finally, rating explanations.}
\end{figure*}

\textbf{Routine Reminders.} In this context, we brainstormed scenarios that captured older adults’ recurring routines and schedules. These scenarios were low-risk and non-urgent, often recurring in older adults’ lives on a daily or regular basis. The primary interaction goal in this context is for the AI to support daily functioning by prompting routine actions and monitoring ongoing routines to provide timely reminders. Interactions in this context are designed to prevent disruptions in routine rather than respond to emergencies. Example scenarios include reminders about grocery store visits, physical exercise, or meals.

\textbf{Emergency Alerts.} This context focuses on high-risk, time-sensitive scenarios involving real-time detection of risks within the home. The primary goal of interaction is to deliver alerts that safeguard a user or their household from immediate threats. Here, the AI is expected to provide rapid alerts to older adults after detecting critical situations. Scenarios in this context prioritized urgency, enabling users to assess risk, understand the AI’s reasoning, and take appropriate action to mitigate harm. Example scenarios include fire hazards, health incidents, security breaches, etc. 

\begin{table*}
    \centering
    \caption{This is a sample set of explanations used in the study to evoke discussions for two scenarios: Scenario 1, where the AI delivers a routine reminder to do some light stretches during the day, and Scenario 2, where the AI delivers an alert to check for smoke in the kitchen for a potential fire hazard.}
    \begin{tabular}{| p{2.25cm} | p{6cm} | p{6cm} |}
    \hline 
      \centering\textbf{Information type in explanations (based on \cite{mathur2024categorizing})} & \centering\textbf{Scenario 1 (Routine Reminder Context) \newline \textit{“I’m reminding you to stretch because...”}} & \textbf{Scenario 2 (Emergency Alert Context) \newline \textit{“I’m alerting you to check the kitchen for smoke because...”}} \\ \hline
      Conversational History & \textit{“..you shared with me yesterday that you were feeling sore, and that light stretching on slow days helps you feel more energized, so I want to make sure you feel good.”} & \textit{“..you shared with me last week that you got distracted while cooking and forgot to switch the stove off, so I want to make sure that does not happen again.”} \\ \hline
      Environmental Data & \textit{“..your smartwatch shows that you’ve been sitting on the couch for more than an hour.”} & \textit{“..the smoke detector in the kitchen just went off, and the temperature is higher than normal.”} \\ \hline
      Activity-Based Inferences & \textit{“..since you haven’t moved around all morning, it might be a good time to take a break and stretch.”}  & \textit{“..since you came to the laundry room while your food was still cooking, the smoke may be coming from the food.”} \\ \hline
      Internal System Logic & \textit{“..I’m about 85\% confident that it’s a good time for some light stretching.”} & \textit{“..I’m about 95\% confident that I detected smoke in the kitchen.”} \\ \hline
    \end{tabular}
\end{table*}

\begin{table*}[t]
\centering
\caption{\textcolor{black}{Participant Overview. Partner A and B denote members of a dyad. For prior AI/LLM use, Rare = once/twice; Occasional = 1-2 times/month; Regular = weekly; Frequently = daily. The randomized explanation order was the same for both contexts for each participant dyad, where 1 = conversational history; 2 = environmental data; 3 = activity-based inferences; 4 = internal system logic.}}
\label{tab:participants}
\renewcommand{\arraystretch}{1.2}
\resizebox{\textwidth}{!}{%
\begin{tabular}{llcccccccccc}
\toprule
\textcolor{black}{Dyad ID} & \textcolor{black}{Partner} & \textcolor{black}{Age} & \textcolor{black}{Gender} & \textcolor{black}{Race/Ethnicity} & \textcolor{black}{Education} & \textcolor{black}{Prior AI/LLM Use} & \textcolor{black}{MCI Status (Y/N)} & \textcolor{black}{Interview Format} & \textcolor{black}{Explanation Order} \\
\midrule

\multirow{2}{*}{\textcolor{black}{P01}} & \textcolor{black}{A} & \textcolor{black}{74} & \textcolor{black}{M} & \textcolor{black}{White} & \textcolor{black}{Bachelor's Degree} & \textcolor{black}{Regular}  & \textcolor{black}{Y} & \textcolor{black}{Dyad} & \textcolor{black}{1 → 2 → 3 → 4} &  \\
                     & \textcolor{black}{B} & \textcolor{black}{72} & \textcolor{black}{F} & \textcolor{black}{Asian} & \textcolor{black}{Bachelor's Degree} & \textcolor{black}{Regular}  & \textcolor{black}{N} & \textcolor{black}{Dyad} &   \\[1ex]

\multirow{2}{*}{\textcolor{black}{P02}} & \textcolor{black}{A} & \textcolor{black}{78} & \textcolor{black}{M} & \textcolor{black}{White} & \textcolor{black}{Bachelor's Degree} & \textcolor{black}{Occasional} & \textcolor{black}{Y} & \textcolor{black}{Dyad} & \textcolor{black}{2 → 3 → 4 → 1} &  \\
                     & \textcolor{black}{B} & \textcolor{black}{70} & \textcolor{black}{F} & \textcolor{black}{White} & \textcolor{black}{Bachelor's Degree} & \textcolor{black}{Rare} & \textcolor{black}{N} & \textcolor{black}{Dyad} & \\[1ex]

\multirow{2}{*}{\textcolor{black}{P03}} & \textcolor{black}{A} & \textcolor{black}{68} & \textcolor{black}{F} & \textcolor{black}{African American} & \textcolor{black}{High School Diploma} & \textcolor{black}{Regular}  & \textcolor{black}{Y} & \textcolor{black}{Dyad} & \textcolor{black}{3 → 4 → 1 → 2} &  \\
                     & \textcolor{black}{B} & \textcolor{black}{71} & \textcolor{black}{M} & \textcolor{black}{African American} & \textcolor{black}{Bachelor's Degree} & \textcolor{black}{Regular} & \textcolor{black}{N} & \textcolor{black}{Dyad} & \\[1ex]

\multirow{2}{*}{\textcolor{black}{P04}} & \textcolor{black}{A} & \textcolor{black}{72} & \textcolor{black}{M} & \textcolor{black}{White} & \textcolor{black}{Graduate Degree} & \textcolor{black}{Occasional} & \textcolor{black}{Y} & \textcolor{black}{Dyad} & \textcolor{black}{4 → 1 → 2 → 3} &  \\
                     & \textcolor{black}{B} & \textcolor{black}{74} & \textcolor{black}{F} & \textcolor{black}{White} & \textcolor{black}{Bachelor's Degree} & \textcolor{black}{Occasional} & \textcolor{black}{N} & \textcolor{black}{Dyad} &  \\[1ex]

\multirow{2}{*}{\textcolor{black}{P05}} & \textcolor{black}{A} & \textcolor{black}{77} & \textcolor{black}{M} & \textcolor{black}{White} & \textcolor{black}{Bachelor's Degree} & \textcolor{black}{Rare} & \textcolor{black}{Y} & \textcolor{black}{Dyad} & \textcolor{black}{1 → 2 → 3 → 4} &  \\
                     & \textcolor{black}{B} & \textcolor{black}{74} & \textcolor{black}{F} & \textcolor{black}{African American} & \textcolor{black}{Bachelor's Degree} & \textcolor{black}{Rare} & \textcolor{black}{N} & \textcolor{black}{Dyad} &  \\[1ex]

\multirow{2}{*}{\textcolor{black}{P06}} & \textcolor{black}{A} & \textcolor{black}{76} & \textcolor{black}{M} & \textcolor{black}{White} & \textcolor{black}{Bachelor's Degree} & \textcolor{black}{Frequent} & \textcolor{black}{Y} & \textcolor{black}{Dyad} & \textcolor{black}{2 → 3 → 4 → 1} &  \\
                     & \textcolor{black}{B} & \textcolor{black}{75} & \textcolor{black}{F} & \textcolor{black}{White} & \textcolor{black}{High School Diploma} & \textcolor{black}{Frequent} & \textcolor{black}{N}& \textcolor{black}{Dyad} &  \\[1ex]

\multirow{2}{*}{\textcolor{black}{P07}} & \textcolor{black}{A} & \textcolor{black}{74} & \textcolor{black}{M} & \textcolor{black}{White} & \textcolor{black}{Bachelor's Degree} & \textcolor{black}{Regular} & \textcolor{black}{Y} & \textcolor{black}{Dyad} & \textcolor{black}{3 → 4 → 1 → 2} &  \\
                     & \textcolor{black}{B} & \textcolor{black}{70} & \textcolor{black}{F} & \textcolor{black}{Asian} & \textcolor{black}{Graduate Degree} & \textcolor{black}{Regular} & \textcolor{black}{N} & \textcolor{black}{Dyad} &  \\[1ex]

\multirow{2}{*}{\textcolor{black}{P08}} & \textcolor{black}{A} & \textcolor{black}{73} & \textcolor{black}{M} & \textcolor{black}{African American} & \textcolor{black}{\textcolor{black}{Graduate Degree}} & \textcolor{black}{None} & \textcolor{black}{Y} & \textcolor{black}{Dyad} & \textcolor{black}{4 → 1 → 2 → 3} &  \\
                     & \textcolor{black}{B} & \textcolor{black}{73} & \textcolor{black}{F} & \textcolor{black}{African American} & \textcolor{black}{Graduate Degree} & \textcolor{black}{Occasional} & \textcolor{black}{N} & \textcolor{black}{Dyad} & \\[1ex]

\multirow{2}{*}{\textcolor{black}{P09}} & \textcolor{black}{A} & \textcolor{black}{67} & \textcolor{black}{M} & \textcolor{black}{White} & \textcolor{black}{Graduate Degree} & \textcolor{black}{Rare} & \textcolor{black}{Y} & \textcolor{black}{Dyad} & \textcolor{black}{1 → 2 → 3 → 4} &  \\
                     & \textcolor{black}{B} & \textcolor{black}{65} & \textcolor{black}{F} & \textcolor{black}{White} & \textcolor{black}{Bachelor's Degree} & \textcolor{black}{Rare} & \textcolor{black}{N} & \textcolor{black}{Dyad} &  \\[1ex]

\multirow{2}{*}{\textcolor{black}{P10}} & \textcolor{black}{A} & \textcolor{black}{69} & \textcolor{black}{F} & \textcolor{black}{Asian} & \textcolor{black}{Graduate Degree} & \textcolor{black}{Regular} & \textcolor{black}{Y} & \textcolor{black}{Dyad} & \textcolor{black}{2 → 3 → 4 → 1} &  \\
                     & \textcolor{black}{B} & \textcolor{black}{73} & \textcolor{black}{M} & \textcolor{black}{Asian} & \textcolor{black}{Graduate Degree} & \textcolor{black}{Regular} & \textcolor{black}{N} & \textcolor{black}{Dyad} & \\[1ex]

\multirow{2}{*}{\textcolor{black}{P11}} & \textcolor{black}{A} & \textcolor{black}{75} & \textcolor{black}{M} & \textcolor{black}{White} & \textcolor{black}{Bachelor's Degree} & \textcolor{black}{Frequent} & \textcolor{black}{Y} & \textcolor{black}{Dyad} & \textcolor{black}{3 → 4 → 1 → 2} &  \\
                     & \textcolor{black}{B} & \textcolor{black}{75} & \textcolor{black}{F} & \textcolor{black}{White} & \textcolor{black}{High School Diploma} & \textcolor{black}{Frequent} & \textcolor{black}{N} & \textcolor{black}{Dyad} & \\[1ex]

\textcolor{black}{P12} & \textcolor{black}{A} & \textcolor{black}{67} & \textcolor{black}{F} & \textcolor{black}{African American} & \textcolor{black}{High School Diploma} & \textcolor{black}{Regular} & \textcolor{black}{Y} & \textcolor{black}{Individual} & \textcolor{black}{4 → 1 → 2 → 3}\\

\bottomrule
\end{tabular}
} 
\end{table*}

\subsubsection{Scenario Ideation and Storyboard Development}  \textbf{Scenario Ideation.} Once the contexts were defined, we then brainstormed a wide range of day-in-the-life scenarios involving older adults and AI systems in the home. We began by generating nearly 50 one-line scenario prompts, grounded in their prevalence in prior research on aging and technology use. We drew inspiration from existing field studies on smart home AI interactions, spanning both Activities of Daily Living (ADLs) and Instrumental Activities of Daily Living (iADLs) (see \cite{pradhan2020use, zubatiy2021empowering, zubatiy2023don, chin2024like, desai2023ok}). Another guiding dimension in our scenario ideation was explanatory depth. Informed by Wolf’s framing of explainability in aging in place contexts \cite{wolf2019explainability}, we prioritized scenarios that offered explanatory depth, especially those where users might question the AI’s behavior due to uncertainty or limited system transparency. Through mutual discussion and agreement among the research team, we selected ten scenarios per context (twenty in total), all of which were developed into storyboards and refined through the piloting process. 

\textbf{Storyboard Development.} We created narrative storyboards that illustrate a visual interaction between older adults and the AI for all scenarios. To anchor participants' understanding, we decided to anthropomorphize the AI as a robot named Rosey (from the popular 60s American sitcom \textit{The Jetsons}). This design choice was made to support mental model calibration for participants and increase engagement by grounding interactions in a playful and familiar metaphor \cite{vandenberghe2016anthropomorphism}. We piloted the initial set of storyboards with five older adult participants to gain insight into the representativeness of the scenarios and whether they were successful in evoking discussions. We refined the scenarios based on pilot feedback, removing those deemed less relevant. The final study included five scenarios per context, each paired with a set of four explanations each. 

The explanations were generated using GPT-4.0 through a prompt describing the study background, context, the four information sources from \cite{mathur2024categorizing}, and were manually verified by a researcher for consistency and clarity. Figure 2 presents sample storyboards used in the study, while Table 1 provides set of explanations for two sample scenarios. The full batch of storyboards and explanations used is included in supplementary files.

\subsection{Participants}
We recruited participants from a clinical program at a healthcare provider. Older adults with a recent diagnosis of Mild Cognitive Impairment (MCI), \textcolor{black}{along with a partner}, are enrolled in this program that focuses on their physical and functional independence. This program provides research opportunities for studying the use of emerging technologies such as AI systems to support older adults in their daily lives. All participants in our study were part of this program, and were either the older adult with MCI, or their co-living partner/spouse. Recruitment requests were sent through IRB-approved channels. The initial screening process was based on comfort in reading and interpreting storyboards and a basic understanding of smart speakers or assistants. We recruited a total of 23 participants: 11 spousal partners\textcolor{black}{/dyads} and one individual older adult, all between the ages of 65 and 78 (M = 72.2; SD = 3.4). 12 participants identified as female, and 11 male. Most participants (96\%) reported having interacted with an AI-driven smart speaker (Home, Alexa, etc.) at least once. \textcolor{black}{We refer to members of each dyad using identifiers (Partner A and Partner B), with Partner A in all cases having MCI. Table 2 provides demographic details, prior AI use, and MCI status for each participant. All study sessions and interviews lasted for around 90 minutes.} Prior to the study, we recruited 5 additional older adults for the piloting stage. During pilots, all older adults expressed \textcolor{black}{their preference to participate} in the study \textcolor{black}{together with their spouse to have mutual support}, particularly when asked to recall details of their \textcolor{black}{shared} daily routines. As a result, \textcolor{black}{all sessions were conducted with both partners present, fostering co-discovery and richer discussions, except for one case (P12) where a partner had to withdraw due to an emergency (noted in Table 2).}  

\begin{figure*}
    \centering
    \includegraphics[width=0.70\linewidth]{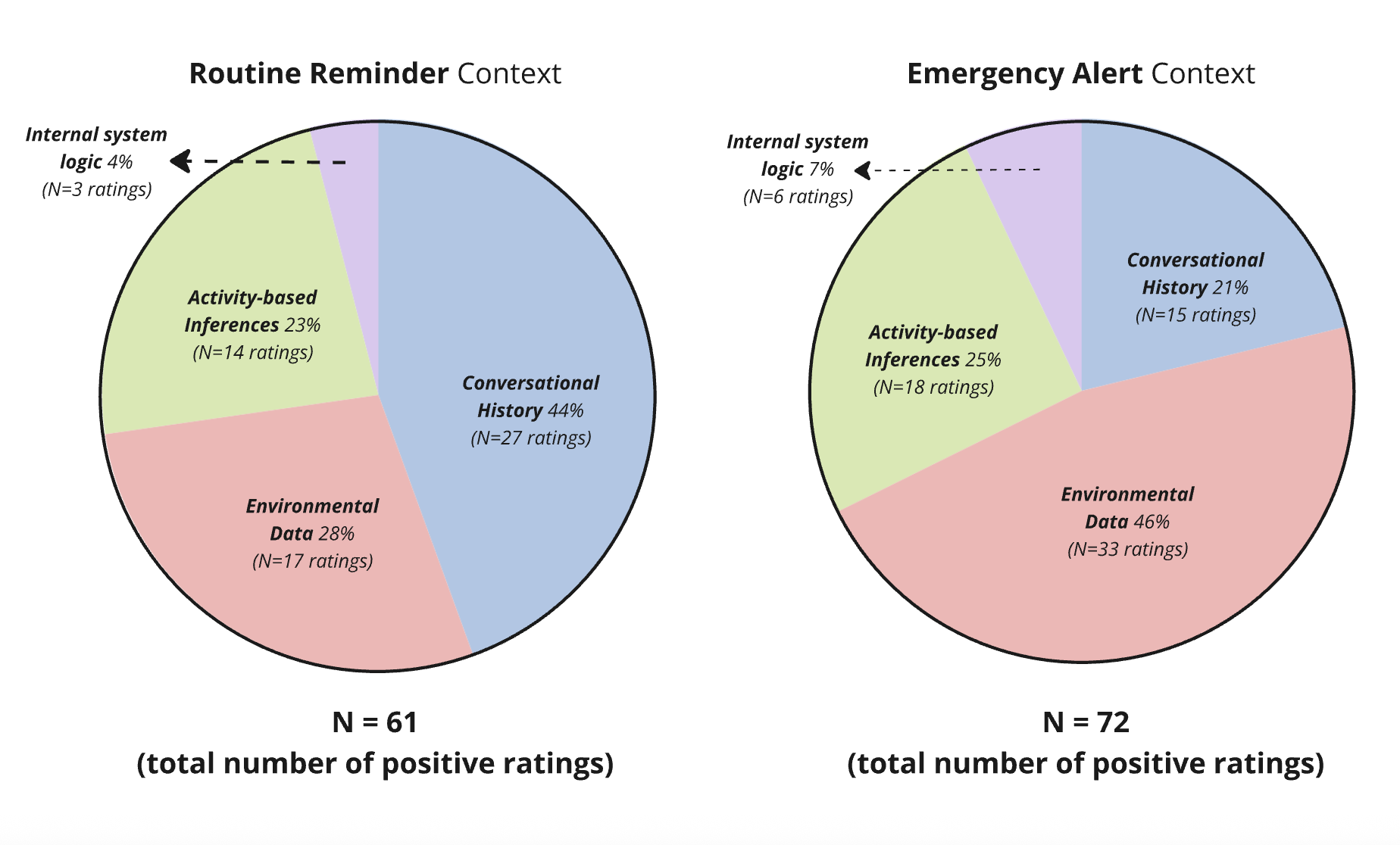}
    \caption{\textcolor{black}{Distribution of positive explanation ratings across the two assistive contexts. \textit{Positive ratings} are the total number of instances when an explanation was rated either \textit{“strongly agree”} or \textit{“agree”} on the Likert-scale, indicating an overall positive-leaning sentiment. These percents \textbf{(rounded up to the nearest whole number)} represent the approximate proportion of positive ratings that each explanation type received. The N in each pie chart is the total number of \textit{positive ratings} for each context and explanation type.}}
    \label{fig:placeholder}
    \Description{This figure shows the distribution of explanation ratings across the two assistive contexts as two pie charts. These percents (rounded up to the nearest decimal) represent the proportion of positive ratings that each explanation type received. Positive ratings are the total number of instances when an explanation was rated either “strongly agree” or “agree” on the Likert-scale, indicating an overall positive-leaning sentiment.}
\end{figure*}

\subsection{Study Procedure}
We conducted the sessions both in-person (in private study rooms with displays) and remotely, based on participant preference. The study session flow is visualized in Figure 3. Participants were first asked to provide consent and complete a demographic survey, and were informed of data recording and usage policies outlined in the IRB. The demographic survey collected their age, cognitive status, care-arrangement, experience with smart assistants, and professional background. Following the survey, we gave them an overview of the process, the two assistive contexts, and the scenarios. To accommodate time and cognitive constraints identified during the pilots, we invited participants to select one scenario from each of the two assistive contexts that resonated most with them. Once participants selected their scenarios, we spent 5-10 minutes discussing rationale for their choices, allowing us to better understand their relevance.

We presented participants with four storyboards for each of the two selected scenarios, with each storyboard featuring an AI explanation drawn from a different information source (e.g., conversational history, environmental data, etc.) in a randomized order. The researcher read each explanation aloud and prompted participants to share their impressions in a semi-structured interview format after each explanation. \textcolor{black}{Interview questions focused on their overall perceptions of the explanations \textbf{(RQ1)}, and variation in those perceptions across the two contexts and source of information within the explanation \textbf{(RQ2)}}. Following the discussion, participants rated the explanation on a 5-point Likert scale \footnote{The Likert scale statement, \textit{"I find this explanation useful"}, asked participants to rate the explanations on 1 (strongly disagree) to 5 (strongly agree).}. The goal of this rating procedure was to contextualize participant perceptions, and \textbf{not} to draw inferential or causal claims. The sessions lasted around 90 minutes, with 3-5 minutes spent sharing each storyboard and probing participants to discuss.

\subsection{Data Analysis}
We transcribed and analyzed all sessions using affinity diagramming, a commonly used method for exploratory research in HCI \cite{davidoff2007rapidly, luria2019re}. We then iteratively grouped and refined emergent themes, with relevant participant quotes used to further develop major themes and subthemes across the sessions. A team of 3 researchers collaboratively discussed the affinity structure, adding subthemes and participant quotes wherever relevant. Disagreements on the affinity structure were resolved through mutual discussion and consensus, and final themes were verified across sessions to describe findings reported in the next section. For the explanation ratings, we aggregated participant ratings for each explanation type across the two contexts and report them as the proportion of positive ratings for each explanation type.

\section{\textcolor{black}{Results and Findings}}

\textcolor{black}{In this section, we first report results from participants' ratings of the different explanation types across the two contexts. Next, we report the in-depth qualitative findings that emerged from our analysis of participant interviews conducted at the end of study sessions.}

\textcolor{black}{During Speed Dating sessions, we asked participants to rate each explanation on a 5-point Likert Scale.} Figure 4 presents the proportion of positive ratings for each explanation type for the two contexts. \textcolor{black}{Here, we define} \textit{positive ratings} as the total number of instances when an explanation was rated either \textit{“strongly agree”} or \textit{“agree”} on the Likert-scale-based statement evaluating their usefulness, indicating an overall positive-leaning sentiment. These percents show how often each explanation type accounted for the total number of positive responses in that context, and highlights which explanation types were rated more favorably relative to others. 

Based on the overall trend, we saw that in the routine reminder context, explanations drawn from conversational history made up the largest proportion of positive ratings, while in the emergency context, explanations from environmental data made up the largest proportion of all the positive ratings. Explanations based on activity-based inferences consistently fell in the mid-range, while internal system logic explanations (expressed through confidence scores) remained the least positively rated in both contexts. Given our limited sample size and exploratory goals, we report only high-level descriptive data to contextualize qualitative findings better and offer a broader view of participant sentiment, and \textbf{do not} intend to make any causal or inferential claims. 

\textcolor{black}{In the following sub-sections, we now discuss the qualitative findings gleaned from the interviews conducted during the Speed Dating sessions.} We identified four overarching themes through analysis of study sessions (N=12) conducted with older adult \textcolor{black}{dyads} (N=23): (1) role of context in explanation preferences, (2) \textcolor{black}{influence of information source in explanations}, (3) explanations for conversational turn-taking, and (4) the multifaceted role of explanations. 
We include representative quotes from Speed Dating sessions to contextualize and support each theme, \textcolor{black}{wherever relevant}. Quotes are labeled using the notation P\#-\textcolor{black}{A/B}, where P\# refers to \textcolor{black}{dyad} ID and \textcolor{black}{A/B} indicates \textcolor{black}{individual within a dyad}. Within participant quotes, all references to \textit{“her”}, \textit{“she”}, or \textit{“Rosey”} are for the AI system, unless stated otherwise.

\subsection{Role of Context in Shaping Explanatory Preferences}

\subsubsection{Routine vs Emergency Contexts}
All participants in our study consistently differentiated their explanation choices and perceptions based on the type of assistive task and the level of risk involved. This led to the finding that explanation requirements are situated and context-dependent. It also demonstrated how participants did not perceive AI’s explanations as isolated and static outputs of the system; instead, they interpreted them as situational communicative responses and evaluated them in relation to their immediate goals, routines, and emotional state.     
We found that for routine, low-risk reminders about their recurring activities (such as meals, grocery visits, pet care, etc.), participants leaned towards explanations that conveyed conversational richness and elaboration, reflecting a \textit{“sense of long-term connection”}. These explanations, often grounded in prior interaction history or user activities, were seen as contributing to a broader sense of companionship with the AI, something which participants felt is \textit{“important if this thing [AI] lives in my home and talks to me all the time”}. Conversationally warm and polite explanations were especially valued in settings where participants anticipated interacting with the AI on a daily basis, imagining it as a long-term presence integrated into their life and accompanying them as their cognitive needs change over time. While reflecting on explanations containing cues from their conversational history (e.g., \textit{Reminding you to stretch because you mentioned feeling sore yesterday..}), \textcolor{black}{P05-B} highlighted her appreciation for conversational warmth in motivating healthy behaviors: 

\begin{quote}
    \textit{“This explanation kind of feels like it has the most empathy. It might sound like a big comparison, but here she kind of sounds like a motherly figure. Like she is saying, ‘Oh I remember this one time you hurt yourself while doing that, so I need to make sure you’re not hurting yourself again in the same way.’ It sounds caring and warm, I like that.”} 
\end{quote}

However, we found that in emergency instances when the AI provided high-risk safety alerts (such as a fall alert, or smoke detection), participants’ preferences shifted. In such cases, they preferred direct, task-focused and concise explanations, rather than elaborate and conversational ones. Regarding conversational verbosity in such situations, several participants expressed how overly verbose or “chatty” explanations may feel counterintuitive. They emphasized that such explanations could unintentionally downplay the urgency of a critical situation, making the AI appear detached from the seriousness of the moment. \textcolor{black}{P04-A} stated: \textit{“If she’s too sweet to me when my kitchen may be on fire, I’d be pretty mad.”} The AI’s role in such situations shifted from being a \textit{socially aware assistant} to a \textit{functional tool} that could enable appropriate decision-making. That is, the primary expectation from the AI in emergencies was to \textit{“just tell me what’s happening and what I need to do.”} In these moments, emotional tone and personalization, while appreciated in routine context, were seen as distractions from the urgent need to act. The elaboration in explanations added to the cognitive load, particularly when participants imagined being already stressed, trying to process the alert. 

Beyond adding to cognitive load, participants also reflected on how overly cautious or polite explanations during emergencies could seem to undermine their sense of competence, particularly when their cognitive abilities might be shifting. \textcolor{black}{P04-B}, reflecting on her husband’s experience, shared:

\begin{quote}
    \textit{“Honestly, with my husband…we’re at a point where he’s starting to forget small things. We had a stove incident a while back, he forgot to switch [stove] off. It affects his sense of self a lot now. I guess most people don’t want to feel like fragile beings who need to be looked after [...] So if she keeps warning him repeatedly and too politely, it might actually make him feel worse.”}
\end{quote}

Interestingly, we also noted that some participants expressed surprise at their own shifting explanation preferences across the two contexts, demonstrating how imagining themselves within each scenario during the Speed Dating sessions surfaced those differences more clearly and intuitively. \textcolor{black}{P02-B} stated:

\begin{quote}
    \textit{“I picked a different one for each [context], right? That’s interesting, I didn’t think I would. But when I imagined myself in those situations, I didn’t want her to talk to me or explain herself to me in the same way for both. Like if she’s reminding me about a healthy habit, she should sound a little friendly. But when she is telling me to check on my husband [for an emergency], good lord I want her to be very direct and sharp about it.”}
\end{quote}

\subsubsection{Emotional State as Cues for Explanations}

Another important dimension of contextual sensitivity that we observed was the influence of participants’ emotional states on how they perceived the explanations. Participants described how the same explanation could feel either supportive or dismissive to them, depending on their mood, stress levels, or emotional state at the time of interaction. These preferences were often linked to the emotional weight of the corresponding scenario, and revealed an important distinction in designing explanations that are either relational or task-oriented. In moments of vulnerability or overwhelm, participants appreciated explanations that acknowledged their emotions and softened the interaction. For example, when discussing how they might respond to a routine AI reminder during an otherwise stressful moment, \textcolor{black}{P07-A} reflected:

\begin{quote}
    \textit{“If I’ve just had a long day or something’s bothering me or just generally if my emotions are running high, I don’t want her throwing a checklist at me. That’s when I need her to talk like a person and not a robot. But if I’m feeling okay and everything’s fine around me, I just want her to get to the point.”}
\end{quote}

This requirement for emotional attunement became particularly salient when participants imagined the AI delivering explanations during emergency situations, where the same explanation could be interpreted differently depending on a user’s response to a crisis. Pointing to a desire for the AI to recognize emotions, participants emphasized that a well-designed system should be able to \textit{“read the room”} and adjust its responses accordingly. \textcolor{black}{P07-A} explained: 

\begin{quote}
    \textit{“The same thing could be read very differently depending on the emotions, right? Like, if someone’s crying in a crisis, you’re not going to give them a hard fact, that’s rude. But if someone’s calm, even in a crisis situation, maybe they just want an action plan. It’s like sometimes a person needs facts, and sometimes they just need a hug.”}
\end{quote}

These reflections speak to the importance of affective and emotional calibration in explanation design. To effectively support older adults, explanations were required to flexibly shift between personal and practical modes depending on emotional and situational cues. Broadly, this recognition of context as a key factor in shaping explanation requirements emphasizes the limitations of applying generalized, \textit{one-size-fits-all} explanatory approaches in AI systems designed to support older adults.  

\subsection{Influence of Information Source on Explanatory Perception}

The explanations in our study were differentiated and designed based on the different sources of information available to an AI in a smart home environment. These sources included a user’s conversational history, environmental data, activity-based inferences, and internal system logic \cite{mathur2024categorizing}. 

We found that participants expressed a preference for explanations that were grounded in personalized and observable information about them, particularly those derived from (1) real-time environmental data and activity-based inferences, and (2) prior conversational history with the AI. While both explanation types were positively received, participants attributed distinct roles to each: explanations grounded in environmental data were associated with more \textit{in-situ} and evidence-based task support, whereas those referencing conversational history helped reinforce the system’s awareness of their routines, preferences, and identity over time. However, some explanations were judged more critically than others. Most participants expressed hesitation towards explanations that relied only on numerical confidence scores without containing any knowledge about them, describing them as \textit{“confusing”}, \textit{“ambiguous”}, and \textit{“difficult to comprehend”}. While the previous finding reported how explanation preferences were shaped by task context, we shift focus in this finding to the explanations themselves, reporting how participants responded to the type of information used to explain the AI's behavior. 

\subsubsection{Explanations Containing Real-Time Environmental Data and Inferences}

Participants consistently favored explanations grounded in real-time environmental data and inferences drawn from them, valuing them for their concreteness and situational relevance. Rather than abstract and generalized reasoning, these explanations drew on observable and tangible inputs, such as temperature readings or heart rate, and were interpreted as indicators of the AI’s situational awareness. Participants frequently described such explanations as \textit{“trustworthy”}, \textit{“factual”}, and \textit{“actionable”}. For many, the presence of a measurable reasoning for the AI’s suggestion positioned it as more grounded, attentive, and responsive to their needs.

In particular, such explanations were valued for the \textit{“confidence”} and \textit{“control”} that participants felt upon knowing what was going on around them, which also prevented them from perceiving the AI’s utterances as sudden or arbitrary. Participants felt more reassured when the explanations were grounded in something tangible, such as a sensor trigger in their environment or data from a wearable device. This reinforced a feeling of staying appropriately informed and active rather than being caught off guard, and made the AI’s behavior feel more intentional to them. 

\begin{quote}
    \textit{“Hearing her say that she alerted me because of the temperature, that makes me feel reassured that she’s paying attention around the house and what’s actually going on around us, and not just guessing. It’s like this is why I’m saying this, here’s the evidence. That makes me feel a little more in control and not like ‘oh god I don’t know what is going on,’ and add to the panic.” \textcolor{black}{(P02-A)}}
\end{quote}

Such explanations contributed to the AI’s perception as a quiet and observant presence in older adults’ homes. Participants described sensor-based cues in explanations as indicators of \textit{“a background presence”}, contributing to enhancing the AI’s usefulness without being overly intrusive, especially in situations where they were independently carrying out tasks, but welcomed help whenever appropriate. For example, \textcolor{black}{(P01-A)} appreciated an explanation drawn from a motion sensor (\textit{“Reminding you to have lunch because you haven’t been in the kitchen all morning”}) by describing it as \textit{“thoughtful but not pushy”}. 

At times, however, the inclusion of sensor triggers and wearable data also invited some questioning, \textcolor{black}{facilitating questions around inaccuracies in explanations}. Some participants were inquisitive about situations when the underlying sensor or wearable data could be inaccurate, and its influence on the explanation. They imagined instances when sensors or wearables could interpret wrong signals, questioning the validity of the explanation’s information source. While discussing a scenario involving an alert to stop exercising for which the AI provided an explanation based on an unusually elevated heart rate, \textcolor{black}{(P08-A)} asked: 

\begin{quote}
    \textit{“It’s helpful if she tells me that my heart rate is very elevated, but what if I’m not really feeling it? [...] Or what if my watch goes fuzzy, which it does many times when I sweat while walking. Will she know when she’s wrong? And what will she do then?”}
\end{quote}

To mitigate this uncertainty, P08-CR further suggested that the AI could conversationally acknowledge her limitations. They suggested that she could say something like, \textit{“I noticed an unusually high heart rate, but your watch has had issues before or I could be wrong, do you feel okay?”}. Notably, this expectation from AI systems to conversationally express uncertainty also speaks to prior work in which expressions of uncertainty (such as \textit{“I’m not sure but it seems like…”}) have shown to appropriately calibrate user confidence \cite{kim2024m, lin2022teaching}.

\subsubsection{Explanations Containing Prior Conversational History}

For explanations that contained references to a user’s prior conversational history with the AI, participants appreciated the AI’s ability to retain and recall relevant interactions and preferences from their previous exchanges. While these explanations were less focused on the immediate situation, they were valued for conveying a sense of memory and familiarity. Unlike sensor-based explanations, which were often described as practical and evidence-based, explanations containing prior conversations were appreciated for their emotional value and the ability to convey the AI’s learning ability, with participants feeling that the AI was \textit{“learning about me over time by recognizing my habits and remembering my choices” }\textcolor{black}{(P08-B)}. This sense of being remembered was particularly meaningful for participants navigating early-stage cognitive changes, offering subtle reinforcement of their routines and sense of continuity.

\begin{quote}
    \textit{“Here it’s like she remembers what we’ve talked about before…that feels nice, like it’s not the first time we’ve spoken. If someday I couldn’t remember small things about me or about him [husband], she would be able to remember them for us and remind us, like little nudges.” \textcolor{black}{(P06-A)}}
\end{quote}

For some participants, \textcolor{black}{particularly those coordinating shared routines and habits}, these explanations offered a form of mutual but gentle accountability. For example, if the AI referenced something previously discussed or agreed upon, participants felt it could reduce conflict or confusion between them by reinforcing consistency across interactions. \textcolor{black}{(P07-B)} shared:

\begin{quote}
    \textit{“Sometimes we can’t remember what we had agreed on, and it’s helpful if Rosey remembers that we said no snacks before dinner, or that he prefers walking in the morning, small things like that, and uses that to explain her reminder to walk for example. Specially if someone has Alzheimer's or a more advanced memory issue, it can snap them back like, ‘ding ding oh yeah I did say that.’ It’s a little thing, but I think it may help to keep us on the same page.”}
\end{quote}

\subsubsection{Skepticism Towards Numbers and Confidence Scores}

While participants generally responded positively to explanations grounded in environmental data and conversational history, their reactions were notably different to explanations containing internal system logic, when expressed through confidence scores. Phrases in explanations such as \textit{“I am 92\% confident...”} were often met with skepticism and confusion, with such explanations getting the lowest ratings. While the ratings gave us a broad overview, our discussions further revealed a deeper nuance that participants’ issue was not just with the numbers themselves, but with how such confidence scores did not include observable information relevant to their lives or environmental context. This ambiguity led to participants questioning both the meaning of the number and the AI’s right to assert confidence without evidence.

\begin{quote}
   \textit{ “If she says she’s 92\% confident, I want to know based on what. What does that even mean in this situation?” \textcolor{black}{(P04-B)}}
\end{quote}

\begin{quote}
    \textit{“She sounds a little overconfident to me. Not that I don’t trust her, but saying it with a number makes her sound like she’s just assuming I’ll go along with it or take her on face value.” \textcolor{black}{(P10-A)}}
\end{quote}

This observation also contradicts a common assumption in XAI that confidence scores in explanations can increase trust calibration by being \textit{“concrete indicators of methodology”} \cite{zhang2020effect, wang2021show}. Our findings revealed that for older adults, explanations with numerical reasoning induced skepticism and warranted further questioning. While some prior research has addressed the risk of \textbf{overtrust} or \textbf{overreliance} in numerical explanations (\cite{binns2018s, yin2019understanding, ehsan2024xai}), our findings pointed towards an alternative reason for the resistance towards such explanations. We found that numerical confidence scores in explanations decreased participants’ willingness to rely on them, i.e., rather than overtrusting or overrelying on confidence scores, older adults were more likely to \textbf{distrust} them. We discuss this further in 6.1.

Some participants linked their inherent skepticism towards numbers to generational differences and limited exposure to AI systems. They emphasized how people in their age bracket tend to value context, reasoning and personal relevance more over abstract metrics and superficial cues of confidence, particularly when delivered by a non-human system.    

\begin{quote}
    \textit{“I think it’s a generational thing right. Our grandkids just talk to these things [AI] like a friend. For us, we want to know what’s going on behind it. And we want it in detail and not just random numbers thrown in.” \textcolor{black}{(P07-B)}}
\end{quote}

Interestingly, a small subset of participants were more neutral or cautiously optimistic about confidence scores, depending on the context and phrasing. Participants with technical professional backgrounds or relatively higher familiarity with technology thought that such scores could add transparency if contextualized appropriately or conveyed in natural language.

\begin{quote}
    \textit{“If you just say something like ‘I thought you might like this,’ that still conveys uncertainty, so I’m not inherently opposed to the numbers. But I think maybe you could bucket the percentages into ranges so it doesn’t sound so clinical. So like anything more than 90\% is very high confidence and so on.” \textcolor{black}{(P05-A)}}
\end{quote}

\begin{quote}
    \textit{“I’m okay with numbers if they’re about something objective, like tracking whether I took my pill or not, when there is a right or wrong answer. But even then, I’d want more than just a number.” \textcolor{black}{(P07-A)}}
\end{quote}

In short, these reactions suggested that the real issue was likely not the presence of AI’s confidence per se, but rather in the way that the confidence was communicated. When presented as a standalone metric, without being anchored in user-relevant context, confidence scores undermined rather than support trust. 

\subsection{Explanations for Conversational Turn-Taking}

A notable theme that emerged \textcolor{black}{regarding perceptions of explanations} was participants’ perception of explanations as dynamic and interactive conversational exchanges, instead of siloed, static AI outputs. They viewed explanations as interactive turn-taking cues that, in most cases, required further negotiation and questioning, and not an end of conversation. When shown a full explanation, participants instead expected it to unfold gradually, aligning with their current needs and emotional state. This style was in contrast to the AI passively delivering the full rationale at once, capable of causing an \textit{“information overload”}. 

This finding also reflected a broader shift in how older adults conceptualized the AI itself, i.e., as a conversational partner that is capable of fluidly responding to questions, mid-sentence interruptions, confusion, or moments of ambiguity, particularly those shaped by cognitive changes. In doing so, participants highlighted an important property of AI explanations that is often overlooked: as scaffolding mechanisms capable of ensuring continued conversational engagement with the AI, rather than a conclusive response meant to serve as a conversational dead-end. Two distinct subthemes emerged from analysis: (1) explanations as part of a multi-turn conversational exchange, and (2) desire for layered and progressive disclosure of information in explanations. 

\subsubsection{Explanations as Multi-Turn Conversational Exchanges}

Participants frequently emphasized their vision for the explanations to mirror “human-like” conversational alignment, in which the explanation would start with an initial statement, and then remain open to a follow-up from the user, allowing them to steer the exchange depending on their cognitive or emotional needs in that moment. Participants also expressed how explanations should provide space to ask further questions and adjust the tone or depth depending on user response. This expectation for a conversational back-and-forth was articulated by \textcolor{black}{(P03-B)}:  

\begin{quote}
    \textit{“So one [explanation] is the follow-up, but the second or third are like the initial ones. So maybe you start with the second, just say here’s what’s going on, this is what it looks like to me, and then you wait for someone to ask for more, based on what they would like to know more about [....] I think the point is I find value in each of these [explanations], so I can’t say I love that one only or hate that [...] I would like to see them all, but at different points.”} 
\end{quote}

As an example for this type of conversational exchange, imagine that a reminder (e.g., \textit{"It’s time for a walk"}) is followed by an initial explanatory justification (e.g., \textit{“..because you’ve been sitting for 90 minutes”}), and then pause to provide the user opportunity to respond, and deliver a follow-up with a softer framing, or respond to user’s request for further clarification. When viewed as a conversational exchange with the AI, participants valued the explanations more because they made the AI feel more \textit{“natural and interactive”} to them. We drew this example from \textcolor{black}{(P03-B)}, who explained:  

\begin{quote}
    \textit{“So initially, you caught my attention, but we’re likely feeling lazy that day, probably why we needed a reminder to walk at all, right? So then, while it could explain with a fact, like hey, you’ve been sitting for too long, or something. But then to soften the blow a little, it could maybe use a gentler explanation after that [...] almost like buttering someone up.”}
\end{quote}

\textcolor{black}{(P03-B)}’s description suggested that explanations may play a dual role: \textit{first}, as an alert or signal that something needs to happen, and \textit{second}, as conversational support that frames the suggestion in an emotionally attuned way. Several participants shared this view, and expressed that they did not want abrupt directives from the AI, instead, they preferred a conversational pacing that allowed them to process the situation and wait to respond on their own terms.

This subtheme also revealed a deeper insight into how older adults assigned accountability to the AI. While reviewing the scenario about an emergency alert, many participants expressed how an explanation that is delivered as a single statement, with no opportunity to follow-up, felt somewhat insufficient. This insufficiency seemed to stem from the fact that it \textit{“closed the door to any further questioning”} and reminded them of \textit{“those horrible chatbots on websites that leave the conversation as soon as you signal frustration”} \textcolor{black}{(P04-B)}. On the other hand, participants perceived the AI’s willingness to remain engaged and to \textit{“stay in the conversation”} as indicative of trustworthiness and accountability, signaling the AI’s openness to feedback.

\subsubsection{Layered and Progressive Disclosure of Information in Explanations}

In addition to multi-turn exchange, participants also emphasized the value of a layered explanation design, one that offers and reveals more details progressively based on participants’ openness or need for additional assurance. In essence, participants expressed a desire to control the preferred level of detail in explanations. We also observed the expectation that the AI triangulate different types of information in delivering explanations, for example, by starting an explanation with information that is easily accessible, and then advancing to more personalized or real-time information, what they called the \textit{“second level of conversational depth”}. \textcolor{black}{(P08-A)} equated this layering of explanations to building a \textit{“conversational flowchart”}, explaining: 

\begin{quote}
    \textit{“It’s definitely the layering aspect, like peeling the layers of a conversation, revealing more details as she goes on. You have to. Like, you’d get kind of a flow chart [...] you’d create a flow chart of all of them. I mean, of everyone she is speaking to, like a unique flowchart for everybody. Like creating a sequence with different layers but for conversation.”} 
\end{quote}

\textcolor{black}{(P08-A)} further expanded on this metaphor of a flowchart by imagining how \textit{“this flowchart would look different for every person, based on what their likes and dislikes are”}, suggesting that they imagined explanations as an inherent part of a longitudinal and adaptable conversational system. Participants also emphasized that different people, or even the same person in different situations, might want to start with different layers of explanations unfolding. For example, some may want to start with brief explanations for routine tasks, while others may prefer layered elaboration from the beginning, hinting at an overall need for triangulation between different types of information that an explanation was capable of conveying. \textcolor{black}{P11-A} put this as the \textit{“working together”} of all the explanations, mentioning that they \textit{“would like to see how she could actually do this. And you can always ask for more, but you don’t want her to assume you want everything every time.”}

\subsection{\textcolor{black}{Multifaceted Roles of Explanations in Everyday Interaction}}

A final theme that emerged from analysis was the different roles or functions that participants attributed to explanations. Variations in participants’ interpretations of explanatory functions underscored the multifaceted nature of explanations, suggesting the need to design them in ways that reflect users’ underlying intentions and distinct goals and motivations. We organize this theme into three high-level explanatory functions described by participants: (1) calibrating urgency, (2) guiding actionability, and (3) providing insights \textcolor{black}{into a partner's well-being}.

\subsubsection{Calibration of Urgency}

A recurring topic of discussion across the sessions was related to what happened after an explanation is delivered, and how it affected participants' decision-making. Participants emphasized on their desire for explanations to help them understand consequences and calibrate the urgency of the AI’s suggestion or alert. They expected the explanations to provide an indication of how quickly they needed to respond, and what the consequences might be if they did not. \textcolor{black}{P08-B} stated: \textit{“I’d like for it to tell me how bad things actually are, kind of like, what’s the worst that could happen if I don’t follow her advice immediately.”} For example, when receiving a potential fall alert about a partner, participants shared that the accompanying explanation mattered significantly in helping them decide whether to rush and respond immediately or take a moment to assess the situation. \textcolor{black}{P06-B} shared a situation where a well-designed explanation could make a big difference:

\begin{quote}
    \textit{“If I get an alert that says, ‘Your husband might have fallen’, I’m already worried. But if it stops there, I’m more worried and don’t know what to do. But if she explains more, and said something like, ‘I heard a loud thud and then no movement for like five minutes,’ that’s when I know I need to drop everything right now [...] like it’s not a false alarm or something.”}
\end{quote}

Participants also mentioned that by contextualizing the stakes of a situation, explanations could help them prioritize between multiple tasks, with \textcolor{black}{P08-B} noting: \textit{“It’s not just the reminder itself, it’s the knowing what happens if I wait [...] that would push me to stop what I’m doing and choose to act immediately.”} 

\subsubsection{Guiding Actionability}
After using the explanation to calibrate a sense of urgency, participants described a subsequent role: guiding their evaluation of the AI’s suggestion and influencing or refining their intention to act on it. Many participants viewed explanations as one of the primary decisive factors in determining the legitimacy of the AI’s suggestion, and in deciding whether or not to act on it. Some indicators that shaped participants’ perception of how compelling the AI’s call to action felt were the information that the explanation conveyed, the explanatory tone and clarity, and whether the AI interrupted them to deliver it. The perceived quality of an explanation was deemed so important to guide actionability that participants expressed how \textit{“a bad explanation would almost make me want to intentionally not follow her suggestion”}, stating:  

\begin{quote}
    \textit{“So she clearly wants me to do this thing, right? Like here she wants me to stop exercising and cool down. The truth is, I probably won’t listen right away [...] so what she says after that matters in whether I will actually stop or not. What more information can she give me about this situation for me to stop right away [...].” \textcolor{black}{(P09-A)}}
\end{quote}

We also noted examples of how explanations could help participants interpret the AI’s recommendations as actionable or dismissible, depending on how articulate and trustworthy the explanation felt. In this context, when participants judged the quality of an explanation as aligning with their conversational preferences and goals, they were more likely to act on it and follow the AI’s suggestion. 

\subsubsection{\textcolor{black}{Explanations as Windows into Partners’ Daily Routines}}

\textcolor{black}{A final important role for explanations emerged in older adults wanting to use AI explanations to stay aware of each other’s routines, especially in households where responsibilities (including aspects of support and care) are naturally shared between couples. As a result of both partners participating jointly, they inadvertently reflected on joint possibilities for making sense of explanations. For many spousal dyads}, explanations served an additional, unique function of providing a window or an overview of their partner’s daily life, preferences and routines. \textcolor{black}{Participants} envisioned AI’s explanations as succinct summaries of their interactions with the AI throughout the day, offering a deeper look into aspects that may be atypical or deviating from their usual routines. In this context, they imagined the explanations to surface details and cues about their partner that may need adjustment or attention. \textcolor{black}{P08-B} articulated this as giving her a sense of “mental peace that she [AI] will be able to flag what’s wrong”, expanding:

\begin{quote}
    \textit{“I don’t need to be hovering all the time, but at the end of the day, if she tells me something like, ‘I reminded Jim\footnote{Name changed for participant privacy.} to eat his food twice today and suggested a break during his walk because his heart rate spiked or something like that..,’ that gives me a good sense of how his day went. It’s like I get a quiet summary of where he might have needed a little nudge [..] and I can step in if something feels off.”}
\end{quote}

Furthermore, they also envisioned explanations as helpful in communicating with extended \textcolor{black}{family members,} such as adult children. Participants imagined how well-designed explanations could benefit family members living at a distance by serving as an overview or a \textit{“window”} into the daily rhythms of their lives, and keep them updated \textcolor{black}{through sharing explanations}. \textcolor{black}{P10-B} reflected on this:

\begin{quote}
    \textit{“I feel like this is the kind of stuff that our daughter would love to know. If Rosey lets her [daughter] know, ‘I reminded mom to have lunch this afternoon because she didn’t go to the kitchen all morning’ or something like that, then she might realize that we’re not eating like we usually do, and could check-in [...] Or if she says, ‘I alerted mom to check the kitchen for smoke twice this week because [....]’, that might tell [daughter] that something’s off.” }
\end{quote}

Related to this reflection, another participant highlighted how \textcolor{black}{they} can also use explanations as a way to \textit{“talk back”} or give feedback to the AI, and to signal what information matters to them and what does not. This use, in turn, could help older adults \textcolor{black}{(and family members)} fine-tune their routines based on \textcolor{black}{recurring} reminders and explanations. \textcolor{black}{This could aid in potentially identifying} behavioral changes, and \textcolor{black}{detecting} early signs of deviation through a longitudinal review of explanatory prompts. \textcolor{black}{P01-B} gave an example of this: 

\begin{quote}
    \textit{“If our son could see these [explanations] too, it would be nice. He often wants to know how we’re doing. I think if he can see or respond to these, it would make him feel more aware. And it could also help Rosey understand what is and isn’t important to us and to him [son] [...]. Like if we could all rate the little explanations every time Rosey explains. For example, if she [AI] determines that the fridge sensor is just not important to us or to our son, she could just tag it as irrelevant for mom.”}
\end{quote}

The ways in which participants articulated the multifaceted functions of explanations highlight the interpretive flexibility with which \textcolor{black}{household members in shared living environments} may engage with explanations. These engagements occur through the lens of their individual roles, responsibilities, and priorities, speaking to the need for explanatory strategies that can adapt to these diverse goals. This perspective on \textcolor{black}{sharing explanations} also raises important considerations around privacy, boundaries, and information disclosure, particularly as older adults may not wish to make all system interactions or details \textcolor{black}{visible} to others, which we discuss in \textcolor{black}{Sections 5.1.1 and 6.} 

\section{Discussion}

In this research, we investigated older adults’ preferences for AI explanations, grounding our understanding in their lived experiences. \textcolor{black}{Now, we discuss some resulting considerations for designing AI explanations for older adults (5.1), and the broader lessons learned from our findings for Human-Centered AI Explanations (5.2).}

\subsection{Design Considerations for AI Explanations for Older Adults}

Our findings underscore that for older adults, the requirements for explainability extend far beyond technical descriptions, and include contextual references to conversational memory, emotional states, and inclusion of a broader sociotechnical environment. Recent work has explored this vision of \textit{“holistic explainability”} through frameworks such as Social Transparency (ST), which outlines the boundaries of what constitutes context in organizational settings \cite{ehsan2021expanding}. Our work extends this perspective into data-rich domestic environments where LLM-based AI systems can draw on multiple sources of information about older adults’ lives to inform explanation design. Consequently, we call for designers to move beyond static, monolithic explanations and toward explanations that support interactive exchanges, emotional attunement, and integration of user context. For example, referencing prior conversations in explanations (e.g., \textit{“..because you mentioned feeling tired yesterday.”}) or modulating explanatory tone in sensitive moments (e.g., \textit{“...because I want to make sure you’re safe.”}) has potential to foster both understanding and trust. Here, we discuss a few other design considerations that emerge from our findings. 

\subsubsection{Explanation Sharing Among Family Members}

We found that older adults often did not consider themselves as the sole recipients of explanations. Many participants imagined explanations being shared with or visible to their children or other \textcolor{black}{family members}. This highlights the potential for explanations to serve as communicative packets in \textcolor{black}{family networks}, and also speaks to prior work on information sharing among \textcolor{black}{informal care networks} and its positive influence \cite{yamashita2018information}. Designers therefore need to consider instances that warrant explanatory visibility to stakeholders beyond primary user(s), and examine how such disclosures can enhance collaborative awareness. \textcolor{black}{In practice, this may include allowing older adults to configure which explanations are shareable, what level of detail is appropriate for different stakeholders, and the conditions under which such sharing occurs.} 

Explanation design must also account for the varying informational needs, expectations, and roles of various stakeholders. \textcolor{black}{This could involve tailoring the informational content and granularity of explanations based on whether they are directed to a co-living partner, an adult child living remotely, or with a member of their clinical team, ensuring that the AI does not disclose more than the older adult intends.} \textcolor{black}{Evidently, when we consider explanations becoming communicative acts,} several questions around social boundaries and autonomy emerge, making it critical to consider appropriate norms and preferences for explanation visibility \cite{czech2023independence, luria2020social}. 

In shared living settings, it is possible that explanations might be overheard, mistimed, or overly revealing. \textcolor{black}{In such cases, designing systems that can modulate explanation modality (such as offering silent summaries, push notifications, or private logs) can help prevent potential inadvertent breaches.} As such, questions around \textit{what} is explained, and more importantly, \textit{when}, to \textit{whom}, and under \textit{what} social conditions necessitate further inquiry. Here, \textcolor{black}{a potential design direction could be} around the recognition and documentation of social roles in the home, such as between co-residents and visitors, or between adult children and older parents, which can be a helpful approach to make the AI more socially aware \cite{luria2020social}. Additionally, recognizing the evolving nature of these roles as care arrangements, cognitive health and family dynamics change over time presents another important nuance for designers to consider \cite{chang2024dynamic}\textcolor{black}{, suggesting opportunities for adaptive explanation routing that evolves alongside household needs.}     

\subsubsection{Incorporating Contextual Elements in Explanation Design}
Our findings provide strong evidence of a shift in older adults' explanation preferences based on task context. While prior work has acknowledged the role of context in AI-mediated interactions, much of it has focused only on emergency situations \cite{zou2025impact, jiang2022chatbot} or within the broader scope of disaster response \cite{zhang2020crowd, chan2023alert}. \textcolor{black}{Our findings highlight the need for designers to explicitly encode the context of use as a factor in explanation design, potentially through rule-based mechanisms, or through context-aware approaches to recognition of user context.} Recent work in recommender systems can offer some valuable directions here. Khaled and Ehab’s ConEX framework for explanations  \cite{khaled2025leveraging} for entertainment-related recommendations develops a taxonomy of user context, combining both static and dynamic user preferences. \textcolor{black}{Inspired by this, designers could consider building similar taxonomy of user preferences through participatory user research, differentiated by what users consider their core explainability values ("static preferences"), and what they consider adaptable values based on the context ("dynamic preferences")}. Interestingly, their system evaluation also revealed that people used explanations to provide feedback about system behavior by rating the explanations \textit{in situ}. Our participants expressed a similar desire to rate or respond to explanations as a form of feedback and to convey their preferences to the AI. \textcolor{black}{In practice, this suggests that explanation design may benefit from incorporating lightweight, interactive feedback channels (e.g., quick verbal confirmations, ratings or corrections) that could enable the AI to adapt its explanatory style over time, and} foster more effective human-AI collaboration.

\subsubsection{Pairing Context With Confidence in Explanations}
Confidence scores, often proposed as straightforward means of communicating algorithmic certainty to users, are a common approach in XAI \cite{moradi2021post, poursabzi2021manipulating, le2023explaining}, particularly when expressed through natural language \cite{li2025conftuner, lin2022teaching}. However, our findings suggest that such representations risk becoming “Explanation Pitfalls” (EPs):  design elements that are perceived as informative but may unintentionally lead to misunderstanding or disengagement \cite{ehsan2024explainability}. EPs commonly occur when users over-estimate the AI’s capability or over-trust explanations, simply because they \textit{“appear trustworthy”} \cite{el2022biases}. Our work with older adults adds another important dimension to EPs: \textbf{distrust} or \textbf{under-trust}, i.e., instances when users question explanations due to lack of contextual grounding or observable justification. Older adults responded with skepticism to phrases like \textit{“I’m 92\% confident..”} particularly when they did not contain any reasoning. This skepticism was not about numbers themselves, but the absence of context: why is the system confident, and based on what? Moving forward, we suggest augmenting numerical explanations with contextual cues to make them more actionable and trustworthy. An example of such an approach is using counterfactuals, or \textit{“what-if”} explanations of confidence scores that have shown to calibrate trust effectively \cite{le2023explaining}.

Another important takeaway from our work is the active role that older adults undertake in negotiating with explanations, challenging their characterization as \textit{passive recipients of technology}, and as users who adopt technological tools out of necessity, rather than agency \cite{morris2000age, desai2022experiential}. This stance often causes devaluation of older adults’ role in the formative stages of AI design \cite{so2024they, pradhan2025understanding}, and leads to underestimation of their needs around AI transparency, accountability and explainability. Our work provides evidence that older adults actively questioned and engaged with explanations in ways that were often in alignment with what younger users may demand from AI systems \cite{binns2018s}. Participants with relatively higher technical familiarity engaged in discussions around explanatory depth. For instance, \textcolor{black}{P08-A}’s description of \textit{“conversational flowcharts”}, \textcolor{black}{highlights that effective explanation design for older adults necessitates considerations that can also inform broader lessons for the design of human-centered AI explanations, which we discuss next.}

\subsection{\textcolor{black}{Broader Lessons for Human-Centered AI Explanations}}

\textcolor{black}{In this section, we take a step back and reflect on lessons from our study that speak to broader conversations on human-centered AI explanations. Grounded in our findings, we discuss research directions towards designing context-sensitive and contextually aware explanations. In line with HCXAI principles that emphasize context, user goals and critical reflection over \textit{one-size-fits-all} transparency \cite{ehsan2020human, kim2024human}, we outline three takeaways from our work: the need for context-sensitive explanations, the role of shared explanations in multi-user settings, and the influence of conversational framing and emotional attunement in explanations.}

\textcolor{black}{\textit{First}, our findings underscore the need for context-sensitive explanations, and illustrate that} while optimizing for technical transparency remains important, equal emphasis must be placed on how explanations are flexibly interpreted within situational context. \textcolor{black}{In our findings, we observed that across both routine and emergency contexts, participants evaluated explanations through the lens of perceived risk, time criticality and their emotional state. In routine contexts, they preferred elaborative, conversational explanations that invited reflection and helped them notice patterns in their habits. In emergency contexts, they favored concise, actionable explanations grounded in concrete evidence. Notably, these shifts in preference were often reflexive rather than deliberate, and yet they consistently shaped whether participants perceived the explanations as helpful or overwhelming. In line with Liao et al.’s argument to move beyond static notions of technical transparency in explanations \cite{liao2021human}, our findings underscore that the value of an explanation also depends on when it appears and how well it aligns with situational demands than on technical accuracy alone.}

\textcolor{black}{\textit{Second}, our findings touch upon an important albeit underexplored dimension of explanation design, i.e., the collaborative nature of technological interactions in multi-user settings. Although our focus was on older adults’ individual perceptions, the shared household contexts in which our participants lived revealed important nuances for designing explanations in multi-user settings. Several participants imagined explanations being shared with partners or adult children. For co-living couples who jointly manage routines, explanations were described as “quiet summaries” that could surface deviations in daily patterns, provide reassurance or prompt check-ins. This reflection, partly in alignment with ongoing work on designing role-appropriate and multi-stakeholder explanatory frameworks \cite{kreiterling2025design, shin2025enhancing}, suggests that in multi-user environments, explanations extend beyond a single user. This recognition, in turn, opens up compelling directions for human-centered explanation design, highlighting the need to study how explanations are shared across people, roles and relationships.}

\textit{Finally}, our findings surfaced several underexplored dimensions of explanation design related to tone, timing, verbosity and conversational framing. Participants frequently described explanations as multi-turn exchanges that could be layered or progressively disclosed, rather than one-shot responses. \textcolor{black}{They also envisioned systems capable of adjusting tone based on the emotional nuances of a situation, characterizing explanations as evolving in response to immediate needs and context.} In this context, as LLM-based systems grow more conversationally expressive, we posit that questions about how explanations are phrased will become increasingly critical to their adoption. \textcolor{black}{Additionally, the prospect of emotionally attuned AI explanations further raises several important implications. While emotionally attuned explanations  (i.e., those that adjust tone, pacing, or expressiveness to the situation) were desired by participants, they carry both appeal and potential risks. While emotional sensitivity can enhance user experience in low-stakes scenarios, overly friendly or confident phrasing may obscure uncertainty or hinder critical reflection from users. This, in turn, can lead to a lack of “healthy skepticism” towards explanations, something which Ehsan et al. present as a foundational design principle for HCXAI \cite{ehsan2020human}. The key takeaway here is the need for an increased focus on when emotional attunement can meaningfully aid understanding and when it risks misleading users, underscoring the importance of explanation strategies that balance tone with informational clarity.} 

\section{\textcolor{black}{Reflecting on Ethical and Privacy Considerations in AI Explanations}}
Several aspects of our findings, particularly those around personalization and use of real-time data, raise important questions about data use, and system memory, and we find it important to reflect on these concerns. Balancing privacy considerations with benefits of personalization requires understanding domain-specific privacy needs and data sharing expectations \cite{huang2025designing}. In our study, we found that older adults’ primary concerns around privacy stem from uncertainty around data use and transmission. While we did not encounter outright opposition to data use in explanations, many participants did share a desire for greater visibility into the data pipeline, including questions of who can access data, and what determines the inclusion (or exclusion) of data in AI-generated explanations. A participant in our study suggested explicitly defining \textit{“no-go zones”} for the AI, for example, excluding any references to financial information when generating explanations. These concerns suggest the need for explainable AI systems to be paired with \textit{“explanations of the explanation”}, i.e., brief disclosures clarifying how the AI uses data to inform its explanation. Brunotte et al.’s work on \textit{“privacy explanations”} offers a useful model for such prompts, showing that they can significantly enhance users’ perception of transparency \cite{brunotte2023privacy}. As LLMs become more capable of using prior interactions to personalize explanations, these explanations may reveal more information about users than they expected the AI to have known, raising new questions around boundaries of acceptable data use in explanations of AI behavior. These realizations point towards the need for rethinking privacy-centered user studies in the LLM era, with focus on exploring how to pair these systems with clearer disclosures that can enable personalization, while also maintaining trust and transparency.

\section{Limitations and Future Work}
Our study primarily involved older adults in early stages of cognitive change; as such, \textcolor{black}{we did not observe strong differences in explanation preferences that could be attributed solely to one partner’s MCI, nor was our study designed for such interpretation. However, our study design can aid future work in examining how cognitive changes may shape explanatory needs.} \textcolor{black}{While the Speed Dating methodology enabled broad exploration of explanation types, its low-fidelity nature inevitably limits ecological realism. Participants were responding to speculative scenarios rather than interacting with a functioning AI system; thus, our study did not capture breakdowns that arise when explanations are incorrect, poorly timed, or technically imperfect. However, this opens up avenues for future work that can build on the insights reported here by developing higher-fidelity interactive prototypes that allow participants to experience explanation uncertainty and error handling in real time, examining how explanation preferences evolve in response to imperfect or ambiguous AI behavior.} In our study, we observed variation in preferences around explanatory tone, content, and timing, and plan to pursue additional research on understanding these explanatory nuances in depth, and explore if and how these preferences vary. Additionally, explanations in our study were grounded in a context where the AI system had access to various information sources, operationalized through speculative storyboard-based vignettes. However, to generalize more broadly, future work should investigate other explanatory methods and techniques (e.g., contrastive, counterfactuals, part-based, etc.), across a wider range of settings and interaction modalities. Finally, our work raises important questions around how explanatory requirements shift as older adults’ routines, environments, and cognitive conditions evolve. Given this dynamic range, designing explanation systems that adapt over time represents another promising direction for future work. 

\section{Conclusion}
As older adults increasingly engage with Conversational AI systems for assistive support while aging in place, it is crucial to design explainable systems that reflect their needs for usability, usefulness, transparency and accountability. In this paper, we take a first step by exploring their preferences for AI explanations across two everyday assistive contexts. We found that explanatory preferences were highly context-dependent and shaped by task risk and participants’ emotional state. We also identified opportunities to design explanations based on personalized information available to the AI, such as prior conversational history or real-time data. Participants in our study also recognized additional avenues for future work, including exploring conversational properties of explanations, adapting explanations to diverse stakeholders and sharing explanations with other \textcolor{black}{family members}.

\begin{acks}
   This material is based upon work supported by the National AI Research Institutes program, supported by the National Science Foundation (NSF) in partnership under Award No. 2112633. Any opinions, findings, and conclusions or recommendations expressed in this material are those of the author(s) and do not necessarily reflect the views of the NSF. We would also like to acknowledge StoryboardThat.com for providing the platform to create the storyboards used in this work.  
\end{acks}

\bibliographystyle{ACM-Reference-Format}
\bibliography{cites}

\end{document}